\documentclass[12pt]{article}
\usepackage[utf8]{inputenc}
\usepackage[dvips]{graphicx}
\usepackage{epsfig}
\usepackage{ulem}
\usepackage{latexsym,amsmath,amsfonts,amssymb}
\usepackage[T1]{fontenc}
\usepackage{rotating}
\usepackage[american]{babel}
\usepackage{bbm}
\usepackage{color}
\usepackage{slashed}
\usepackage[unicode]{hyperref}
\usepackage{lscape}
\usepackage{bigints}
\usepackage{enumerate}
\usepackage[shortlabels]{enumitem}
\newtheorem{lemma}{Observation}
\usepackage{tikz}
\usetikzlibrary{decorations.pathmorphing}
\usetikzlibrary{decorations.pathreplacing}
\usetikzlibrary{arrows.meta}
\tikzset{
  >={To[length=5pt]}
  }
\usetikzlibrary{shapes, shapes.geometric, shapes.symbols, shapes.arrows, shapes.multipart, shapes.callouts, shapes.misc}
\tikzset{snake it/.style={decorate, decoration=snake}}
\tikzset{7brane/.style={circle, draw=black, fill=black,ultra thick,inner sep=1.5 pt, minimum size=1 pt,}, c/.default={4pt}}
\tikzset{cross/.style={cross out, draw=black,thick, minimum size=2*(#1-\pgflinewidth), inner sep=0pt, outer sep=0pt}, cross/.default={5pt}}
\tikzset{big7brane/.style={circle, draw=black, fill=black,ultra thick,inner sep=2.5 pt, minimum size=1 pt,}, c/.default={4pt}}
\tikzset{u/.style={circle, draw=black, fill=white,inner sep=2 pt, minimum size=2 pt,},f/.style={square, draw=black, fill=white,ultra thick,inner sep=4 pt, minimum size=2 pt,}}
\tikzset{so/.style={circle, draw=black, fill=red,inner sep=2 pt, minimum size=2 pt,},f/.style={square, draw=black, fill=white,ultra thick,inner sep=4 pt, minimum size=2 pt,}}
\tikzset{sp/.style={circle, draw=black, fill=blue,inner sep=2 pt, minimum size=2 pt,},f/.style={square, draw=black, fill=white,ultra thick,inner sep=4 pt, minimum size=2 pt,}}
\tikzset{uf/.style={rectangle, draw=black, fill=white,inner sep=3 pt, minimum size=4 pt,}}
\tikzset{spf/.style={rectangle, draw=black, fill=blue, thick,inner sep=3 pt, minimum size=4 pt, circle, draw=black, fill=blue,thick,inner sep=2 pt, minimum size=2 pt,},f/.style={square, draw=black, fill=white,ultra thick,inner sep=4 pt, minimum size=2 pt,}}
\tikzset{sof/.style={rectangle, draw=black, fill=red, thick,inner sep=3 pt, minimum size=4 pt,}}
\usetikzlibrary{positioning}
\usetikzlibrary{arrows}
\usetikzlibrary{decorations.pathreplacing}
\usetikzlibrary{shapes}
\def\im{Invent. Math.}

\def\hat{\widehat}
\def\a{\alpha}
\def\b{\beta}
\def\c{\gamma}
\def\d{\delta}
\def\f{\phi}               
\def\vf{\varphi}  
\def\tvf{\tilde{\varphi}}
\def\vp{\varphi}
\def\g{\gamma}
\def\h{\eta}
\def\j{\psi}
\def\k{\kappa}                    
\def\l{\lambda}
\def\m{\mu}
\def\n{\nu}
\def\o{\omega}  \def\w{\omega}
\def\p{\pi}
\def\q{\theta}  \def\th{\theta}                  
\def\r{\rho}                                     
\def\s{\sigma}                                   
\def\t{\tau}
\def\u{\upsilon}
\def\x{\xi}
\def\z{\zeta}
\def\pt{\tilde{\varphi}}
\def\tt{\tilde{\theta}}
\def\lab{\label}
\def\6{\partial}
\def\wg{\wedge}
\def\bpsi{\bar{\psi}}
\def\bt{\bar{\theta}}
\def\bvf{\bar{\varphi}}

\DeclareMathOperator{\tr}{tr}

\newcommand{\be}{\begin{equation}}
\newcommand{\ee}{\end{equation}}
\newcommand{\beq}{\begin{equation}}
\newcommand{\eeq}{\end{equation}}
\newcommand{\bea}{\begin{eqnarray}}
\newcommand{\eea}{\end{eqnarray}}

\newcommand{\ba}{\begin{eqnarray}}
\newcommand{\ea}{\end{eqnarray}}

\newcommand{\beqs}{\begin{eqnarray}}
\newcommand{\eeqs}{\end{eqnarray}}
\newcommand{\bal}{\begin{aligned}}
\newcommand{\eal}{\end{aligned}}
\makeatletter
\newcommand\setItemnumber[1]{\setcounter{enum\romannumeral\@enumdepth}{\numexpr#1-1\relax}}
\makeatother
\begin{document}
\baselineskip=15.5pt
\pagestyle{plain}
\setcounter{page}{1}

\def\del{{\partial}}
\def\vev#1{\left\langle #1 \right\rangle}
\def\cn{{\cal N}}
\def\co{{\cal O}}


\def\IC{{\mathbb C}}
\def\IR{{\mathbb R}}
\def\IZ{{\mathbb Z}}
\def\RP{{\bf RP}}
\def\CP{{\bf CP}}
\def\Poincaré{{Poincar\'e }}
\def\tr{{\rm tr}}
\def\tp{{\tilde \Phi}}

\def\TL{\hfil$\displaystyle{##}$}
\def\TR{$\displaystyle{{}##}$\hfil}
\def\TC{\hfil$\displaystyle{##}$\hfil}
\def\TT{\hbox{##}}
\def\HLINE{\noalign{\vskip1\jot}\hline\noalign{\vskip1\jot}}
\def\seqalign#1#2{\vcenter{\openup1\jot
   \halign{\strut #1\cr #2 \cr}}}
\def\lbldef#1#2{\expandafter\gdef\csname #1\endcsname {#2}}
\def\eqn#1#2{\lbldef{#1}{(\ref{#1})}%
\begin{equation} #2 \label{#1} \end{equation}}
\def\eqalign#1{\vcenter{\openup1\jot
     \halign{\strut\span\TL & \span\TR\cr #1 \cr
    }}}

\def\eno#1{(\ref{#1})}
\def\href#1#2{#2}
\def\half{\frac{1}{2}}



\def\ads{{\it AdS}}
\def\adsp{{\it AdS}$_{p+2}$}
\def\cft{{\it CFT}}

\newcommand{\ber}{\begin{eqnarray}}
\newcommand{\eer}{\end{eqnarray}}

\newcommand{\beqar}{\begin{eqnarray}}
\newcommand{\cN}{{\cal N}}
\newcommand{\cO}{{\cal O}}
\newcommand{\cA}{{\cal A}}
\newcommand{\cT}{{\cal T}}
\newcommand{\cF}{{\cal F}}
\newcommand{\cC}{{\cal C}}
\newcommand{\cR}{{\cal R}}
\newcommand{\cW}{{\cal W}}
\newcommand{\eeqar}{\end{eqnarray}}
\newcommand{\tht}{\thteta}
\newcommand{\lm}{\lambda}\newcommand{\Lm}{\Lambda}


\newcommand{\nonu}{\nonumber}
\newcommand{\oh}{\displaystyle{\frac{1}{2}}}
\newcommand{\dsl}
   {\kern.06em\hbox{\raise.15ex\hbox{$/$}\kern-.56em\hbox{$\partial$}}}
\newcommand{\id}{i\!\!\not\!\partial}
\newcommand{\as}{\not\!\! A}
\newcommand{\ps}{\not\! p}
\newcommand{\ks}{\not\! k}
\newcommand{\D}{{\cal{D}}}
\newcommand{\dv}{d^2x}
\newcommand{\Z}{{\cal Z}}
\newcommand{\N}{{\cal N}}
\newcommand{\Dsl}{\not\!\! D}
\newcommand{\Bsl}{\not\!\! B}
\newcommand{\Psl}{\not\!\! P}

\newcommand{\eeqarr}{\end{eqnarray}}
\newcommand{\ZZ}{{\rm \kern 0.275em Z \kern -0.92em Z}\;}


\def\del{{\delta^{\hbox{\sevenrm B}}}} \def\ex{{\hbox{\rm e}}}
\def\azb{A_{\bar z}} \def\az{A_z} \def\bzb{B_{\bar z}} \def\bz{B_z}
\def\czb{C_{\bar z}} \def\cz{C_z} \def\dzb{D_{\bar z}} \def\dz{D_z}
\def\im{{\hbox{\rm Im}}} \def\mod{{\hbox{\rm mod}}} \def\tr{{\hbox{\rm Tr}}}
\def\ch{{\hbox{\rm ch}}} \def\imp{{\hbox{\sevenrm Im}}}
\def\trp{{\hbox{\sevenrm Tr}}} \def\vol{{\hbox{\rm Vol}}}
\def\rl{\Lambda_{\hbox{\sevenrm R}}} \def\wl{\Lambda_{\hbox{\sevenrm W}}}
\def\fc{{\cal F}_{k+\cox}} \def\vev{vacuum expectation value}
\def\nodiv{\mid{\hbox{\hskip-7.8pt/}}}
\def\ie{{\em i.e.}}
\def\ie{\hbox{\it i.e.}}

\def\CC{{\mathchoice
{\rm C\mkern-8mu\vrule height1.45ex depth-.05ex
width.05em\mkern9mu\kern-.05em}
{\rm C\mkern-8mu\vrule height1.45ex depth-.05ex
width.05em\mkern9mu\kern-.05em}
{\rm C\mkern-8mu\vrule height1ex depth-.07ex
width.035em\mkern9mu\kern-.035em}
{\rm C\mkern-8mu\vrule height.65ex depth-.1ex
width.025em\mkern8mu\kern-.025em}}}

\def\RR{{\rm I\kern-1.6pt {\rm R}}}
\def\NN{{\rm I\!N}}
\def\ZZ{{\rm Z}\kern-3.8pt {\rm Z} \kern2pt}
\def\IB{\relax{\rm I\kern-.18em B}}
\def\ID{\relax{\rm I\kern-.18em D}}
\def\II{\relax{\rm I\kern-.18em I}}
\def\IP{\relax{\rm I\kern-.18em P}}
\newcommand{\CS}{{\scriptstyle {\rm CS}}}
\newcommand{\CSs}{{\scriptscriptstyle {\rm CS}}}
\newcommand{\rc}{\nonumber\\}
\newcommand{\bear}{\begin{eqnarray}}
\newcommand{\eear}{\end{eqnarray}}

\newcommand{\LL}{{\cal L}}

\def\mani{{\cal M}}
\def\calo{{\cal O}}
\def\calb{{\cal B}}
\def\calw{{\cal W}}
\def\calz{{\cal Z}}
\def\cald{{\cal D}}
\def\calc{{\cal C}}

\def\to{\rightarrow}
\def\ele{{\hbox{\sevenrm L}}}
\def\ere{{\hbox{\sevenrm R}}}
\def\zb{{\bar z}}
\def\wb{{\bar w}}
\def\nodiv{\mid{\hbox{\hskip-7.8pt/}}}
\def\menos{\hbox{\hskip-2.9pt}}
\def\dr{\dot R_}
\def\drr{\dot r_}
\def\ds{\dot s_}
\def\da{\dot A_}
\def\dga{\dot \gamma_}
\def\ga{\gamma_}
\def\dal{\dot\alpha_}
\def\al{\alpha_}
\def\cl{{closed}}
\def\cls{{closing}}
\def\vev{vacuum expectation value}
\def\tr{{\rm Tr}}
\def\to{\rightarrow}
\def\too{\longrightarrow}


\def\a{\alpha}
\def\b{\beta}
\def\c{\gamma}
\def\d{\delta}
\def\e{\epsilon}           
\def\F{\Phi}
\def\f{\phi}               
\def\vf{\varphi}  \def\tvf{\tilde{\varphi}}
\def\vp{\varphi}
\def\g{\gamma}
\def\h{\eta}
\def\j{\psi}
\def\k{\kappa}                    
\def\l{\lambda}
\def\m{\mu}
\def\n{\nu}
\def\o{\omega}  \def\w{\omega}
\def\q{\theta}  \def\th{\theta}                  
\def\r{\rho}                                     
\def\s{\sigma}                                   
\def\t{\tau}
\def\u{\upsilon}
\def\x{\xi}
\def\X{\Xi}
\def\z{\zeta}
\def\pt{\tilde{\varphi}}
\def\tt{\tilde{\theta}}
\def\lab{\label}
\def\6{\partial}
\def\wg{\wedge}
\def\atanh{{\rm arctanh}}
\def\bpsi{\bar{\psi}}
\def\bt{\bar{\theta}}
\def\bvf{\bar{\varphi}}

%



\newfont{\namefont}{cmr10}
\newfont{\addfont}{cmti7 scaled 1440}
\newfont{\boldmathfont}{cmbx10}
\newfont{\headfontb}{cmbx10 scaled 1728}





\newcommand{\re}{\,\mathbb{R}\mbox{e}\,}
\newcommand{\hyph}[1]{$#1$\nobreakdash-\hspace{0pt}}
\providecommand{\abs}[1]{\lvert#1\rvert}
\newcommand{\Nugual}[1]{$\mathcal{N}= #1 $}
\newcommand{\sub}[2]{#1_\text{#2}}
\newcommand{\partfrac}[2]{\frac{\partial #1}{\partial #2}}
\newcommand{\bsp}[1]{\begin{equation} \begin{split} #1 \end{split} \end{equation}}
\newcommand{\calF}{\mathcal{F}}
\newcommand{\calO}{\mathcal{O}}
\newcommand{\calM}{\mathcal{M}}
\newcommand{\calV}{\mathcal{V}}
\newcommand{\bbZ}{\mathbb{Z}}
\newcommand{\bbC}{\mathbb{C}}
\newcommand{\cK}{{\cal K}}

\newcommand{\Thq}{\Theta\left(\r-\r_q\right)}
\newcommand{\Dq}{\d\left(\r-\r_q\right)}
\newcommand{\kten}{\kappa^2_{\left(10\right)}}
\newcommand{\pbi}[1]{\imath^*\left(#1\right)}
\newcommand{\ho}{\hat{\omega}}
\newcommand{\tth}{\tilde{\th}}
\newcommand{\tf}{\tilde{\f}}
\newcommand{\tj}{\tilde{\j}}
\newcommand{\tw}{\tilde{\omega}}
\newcommand{\tz}{\tilde{z}}
\newcommand{\prj}[2]{(\partial_r{#1})(\partial_{\j}{#2})-(\partial_r{#2})(\partial_{\j}{#1})}
\def\atanh{{\rm arctanh}}
\def\sech{{\rm sech}}
\def\csch{{\rm csch}}
\allowdisplaybreaks[1]

\def\red{\textcolor[rgb]{0.98,0.00,0.00}}

\newcommand{\Dan}[1] {{\textcolor{blue}{#1}}}

\numberwithin{equation}{section}

\newcommand{\Tr}{\mbox{Tr}}    


%

\setcounter{footnote}{0}
\renewcommand{\theequation}{{\rm\thesection.\arabic{equation}}}

\begin{titlepage}

\begin{center}

\vskip .5in 
\noindent

{\Large \bf{ Electrostatic description of 3d $\mathcal{N}=4$ linear quivers} }
\bigskip\medskip
\\
Mohammad Akhond\footnote{akhondmohammad@gmail.com}, Andrea Legramandi\footnote{andrea.legramandi@swansea.ac.uk} and Carlos Nunez\footnote{c.nunez@swansea.ac.uk}
\\

\bigskip\medskip
{\small 

Department of Physics, Swansea University, Swansea SA2 8PP, United Kingdom}

\vskip .5cm 
\vskip .9cm 
\abstract{	We present the holographic dual for the strongly coupled, low energy dynamics of   {\it balanced} ${\cal N}=4$ field theories in $(2+1)$ dimensions. The infinite family
of Type IIB backgrounds with AdS$_4\times S^2\times S^2$ factors is described in terms of a Laplace problem with suitable boundary conditions. The system describes an array of D3, NS5 and D5 branes.
We study various aspects of these Hanany--Witten set-ups (number of branes, linking numbers, dimension of the Higgs and Coulomb branches) and encode them in holographic calculations. A generic expression for the Free Energy/Holographic Central Charge is derived. These quantities are then calculated explicitly in various general examples. We also discuss how Mirror Symmetry is encoded in our Type IIB backgrounds. The connection with previous results in the bibliography is made.}
	
\end{center}

\noindent

\noindent
\vskip .5cm
\vskip .5cm
\vfill
\eject

\end{titlepage}

\setcounter{footnote}{0}


\normalsize

%
\renewcommand{\theequation}{{\rm\thesection.\arabic{equation}}}

\tableofcontents

\section{Introduction}
The conjunction of conformal symmetry and supersymmetry proved to be a very powerful tool to analyse the existence and dynamics of fixed points for field theories in dimension $d+1$ (with $d=0,....,5$).
In this line, Maldacena's AdS/CFT conjecture \cite{Maldacena:1997re} played a central role motivating the construction of AdS$_D$ backgrounds in consistent theories of gravity.
\\
For the particular case of half-BPS backgrounds with isometries $SO(2,D-1)\times SU(2)$, great progress was achieved. In fact,  infinite classes  of backgrounds of the form AdS$_D\times S^2\times \Sigma_{8-D}$ have been constructed for the cases $D=2,....,7$. For some  values of $D$, these backgrounds are described in terms of a {\it potential function}. This potential satisfies a Laplace equation which needs of initial and boundary conditions to be well-defined.

It is in these initial or boundary conditions that the healthy-character of the background is encoded and where the connection with the dual CFT is made concrete. Indeed, the presence of a `rank' function (so called as it encodes the ranks of the colour and flavour groups of the field theory) turns out to be the initial condition of the Laplace equation.
For the cases $D=2,3,5,7$ the formalism, backgrounds and dual field theories are  respectively described in the papers 
\cite{Lozano:2020txg}-\cite{Lozano:2021rmk} (for AdS$_2$),  \cite{Couzens:2017way}-\cite{Lozano:2019ywa}
(for AdS$_3$), \cite{Gaiotto:2009gz}-\cite{Nunez:2018qcj} (for AdS$_5$) and  \cite{Apruzzi:2013yva}-\cite{Bergman:2020bvi}  (for AdS$_7$).
The cases $D=4$ and $D=6$ corresponding with SCFTs in dimension three and five have a very elegant formulation in terms of holomorphic functions, but the connection with the dual SCFT is a bit more laborious. See 
\cite{DHoker:2007hhe}-\cite{Coccia:2020wtk} (for AdS$_4$) and \cite{DHoker:2016ujz}-\cite{Uhlemann:2020bek} (for AdS$_6$) respectively. The case $D=6$ was recently written in the `electrostatic' context (Laplace equation and boundary-initial conditions) in \cite{Legramandi:2021uds}.

A first goal of this paper is to complete the picture and write the case of AdS$_4$ in this electrostatic formalism.
In fact, we present a holographic dual formulation of  ${\cal N}=4$, d=3 superconformal field theories (linear quivers),  involving AdS$_4\times S^2\times S^2$ backgrounds in type IIB. The case we deal with in this paper is that of {\it balanced} linear quivers, that is for each gauge node the number of fields transforming in the (bi)fundamental is twice the rank of the gauge group (or  $N_f=2N_c$).
\\
The contents of this work are distributed across the coming sections as follows.

In Section \ref{section2xx} we study the background, the defining Laplace PDE, its initial and boundary conditions. Possible singular behaviours in the spacetime are discussed.

In Section \ref{Pagechargessection} we study the Page charges. Imposing their quantisation determines the range of some coordinates and the character of the rank function (the initial condition for the Laplace equation), that we determine to be a piecewise, continuous, linear function. We study the associated Hanany--Witten set-up and linking numbers, giving a holographic expression for them. We  algorithmically associate a balanced linear quiver with a given supergravity solution. We also find a generic expression for the holographic central charge. This is a purely geometric quantity that counts the number of degrees of freedom of the dual CFT. Usually, this is also identified as proportional to the Free Energy of the CFT when formulated on a three-sphere.

In Section \ref{examplessection} we discuss generic examples of  linear quivers and study all the quantities defined in Section \ref{Pagechargessection}: charges, Hanany--Witten set-ups, linking numbers, central charge. We discuss special limits of our examples and compare them with previously found results.

Section \ref{sectionmirrorxx} summarises some known field theoretical aspects of the dual 3d ${\cal N}=4$ SCFTs, dwelling in particular with Mirror symmetry. We present a purely geometric version of Mirror symmetry, mapping balanced quivers into balanced quivers. This geometric correspondence exchanges between NS5 and D5 branes and the dimensions of the Higgs and Coulomb branches of the theories, all being nicely realised as a simple operation in the string description. As a spin-off, we present a (not-mirror) transformation that maps a balanced linear quiver into a different  one (still balanced and linear), both sharing the same central charge.
Some of the content of this section might illuminate future work and this, together with other possible lines of investigation, are presented in Section \ref{conclusions}, with a summary of the main results obtained in the paper and some concluding remarks. Various extensive and dense appendices complement this paper.

\section{Geometry}\label{section2xx}

We start this section by writing explicitly the infinite family of type IIB supergravity backgrounds we work with. 

We are after solutions dual to ${\cal N}=4$ super-conformal field theories. This implies that the background must have  isometries $\text{SO}(2,3)\times \text{SU}(2)_C\times \text{SU}(2)_H$ and preserve four Poincar\'e supercharges to match the global symmetries of the field theory.
Hence, our geometries must contain an AdS$_4$ factor and a couple of two spheres  $S^2_1(\theta_1,\varphi_1)$ and  $S^2_2(\theta_2,\varphi_2)$. There are two extra directions labelled by $(\sigma,\eta)$.  The  presence of $\text{SO}(2,3)\times \text{SU}(2)_C\times \text{SU}(2)_H$  isometries allow for warp factors that depend only on $(\sigma,\eta)$. The background must have the form
 AdS$_4\times S^2_1\times S^2_2\times \Sigma_2(\sigma,\eta)$.  
The Ramond and Neveu-Schwarz fields must also respect the above mentioned isometries.
 
The preservation of four Poincar\'e supersymmetries implies that the generic type IIB background
can be casted in terms of a function $V(\sigma,\eta)$. 
In string frame the solution reads
\begin{eqnarray}
& & ds_{10,st}^2= f_1(\sigma,\eta)\Big[ds^2(\text{AdS}_4) + f_2(\sigma,\eta) d s^2 (S^2_1)+ f_3(\sigma,\eta) d s^2 (S^2_2)+ f_4(\sigma,\eta)(d\sigma^2+d\eta^2) \Big], \nonumber\\[2mm]
& &e^{-2\Phi}=f_5(\sigma,\eta), \;\; B_2=f_6(\sigma,\eta) \text{Vol}(S^2_1),\;\;C_2= f_7(\sigma,\eta) \text{Vol}(S^2_2),\;\;\; \tilde{C}_4= f_8(\sigma,\eta) \text{Vol(AdS}_4), \nonumber\\ [2mm]
& & f_1=\frac{\pi}{2}\sqrt{\frac{\sigma^3 \partial^2_{\eta \sigma}V}{\partial_{\sigma}(\sigma \partial_{\eta} V)}},\;\; f_2= -\frac{\partial_\eta V \partial_{\sigma}(\sigma \partial_{\eta} V)}{\sigma \Lambda},\;\;f_3= \frac{\partial_{\sigma}(\sigma \partial_{\eta} V)}{\sigma \partial^2_{\eta \sigma} V},\;\; f_4= -\frac{\partial_{\sigma}(\sigma \partial_{\eta} V)}{\sigma^2 \partial_{\eta}V},\nonumber\\[2mm]
& & f_5=-16\frac{\Lambda  \partial_{\eta}V}{ \partial^2_{\eta\sigma} V} , \;\;\;\;  f_6= \frac{\pi}{2} \left(\eta -\frac{\sigma  \partial_{\eta}V \partial_{\eta}^2 V}{\Lambda }\right) ,\;\;\;\;  f_7 = -2 \pi \left(\partial_\sigma (\sigma  V)-\frac{\sigma  \partial_{\eta}V \partial_{\eta}^2 V}{\partial^2_{\eta\sigma} V}\right) , \nonumber\\[2mm]
& & f_8 = -\pi^2 \sigma ^2 \left(3 \partial_{\eta} V+\frac{\sigma  \partial_{\eta} V \partial_{\eta}^2 V}{\partial_\sigma (\sigma  \partial_{\eta}V) }\right), \;\;\;\; \Lambda = \partial_{\eta}V \partial^2_{\eta\sigma} V + \sigma \left(( \partial^2_{\eta\sigma} V )^2 + ( \partial^2_{\eta} V )^2 \right). \label{background}
\end{eqnarray}
Where the fluxes are defined from the potentials as follows,
\begin{equation}
F_1=0 , \quad H_3 = d B_2 \quad F_3 = d C_2, \quad F_5= d \tilde{C}_4 + *d \tilde{C}_4 .
\end{equation}
Using Mathematica, we have checked that the configuration in eq.(\ref{background}) is solution to the Type IIB equations of motion (see Appendix \ref{appendix1}), if the function $V(\sigma,\eta)$ satisfies,
\begin{equation}
\partial_\sigma \left(\sigma^2 \partial_\sigma V\right) +\sigma^2 \partial^2_\eta V=0.\label{diffeq} 
\end{equation}
This infinite family of solutions can be mapped into the backgrounds described by D'Hoker, Estes and Gutperle in \cite{DHoker:2007hhe}. See Appendix \ref{mapDEGK} for the details of this map.

For the backgrounds in eq.(\ref{background}) to be well defined,
$e^{2\Phi}$ and the metric warping functions must be real and positive. For the class of solutions we analyse in the next sections, we assume  the symmetry $V(-\sigma,\eta)= - V(\sigma,\eta)$. Under  the `parity' change $\sigma \to - \sigma$, the quantity
$\Lambda(-\sigma,\eta)=-\Lambda(\sigma,\eta)$. The reader can check that the functions $f_1, \dots, f_5$ are invariant under this `parity' transformation. Hence, the solution for negative $\sigma$ is well defined as long as the one with positive $\sigma$ is. The required positivity condition for the dilaton and warping functions is
\begin{equation}
\label{eq:positivity}
- \sigma \frac{\partial^2_{\eta \sigma} V}{\partial_\eta V } \ge 1 \, .
\end{equation}
As a direct result of the above condition we have
\begin{eqnarray}
& &\sigma \Lambda =  \sigma \partial_{\eta}V \partial^2_{\eta\sigma} V + \left(|\sigma \partial^2_{\eta\sigma} V |^2 + (\sigma  \partial^2_{\eta} V )^2 \right) \ge \sigma \partial_{\eta}V \partial^2_{\eta\sigma} V + |\sigma \partial_{\eta}V \partial^2_{\eta\sigma} V|+ (\sigma \partial^2_{\eta} V )^2 \ge 0 \, , \nonumber \\
& & \sigma \frac{\partial_{\sigma}(\sigma \partial_{\eta} V)}{\sigma  \partial^2_{\eta \sigma}V} = 1 + \frac{ \partial_\eta V}{\sigma  \partial^2_{\eta \sigma}V} \ge 0 \, .
\end{eqnarray}
The positivity of $f_1, f_2,f_3,f_4, f_5$ is derived as a consequence of $\sigma \Lambda \ge 0$ and eq.\eqref{eq:positivity}.
\subsection{Study of the partial differential equation }
Let us now study the partial differential equation (\ref{diffeq}). Define $V(\sigma,\eta)= \frac{\hat{V}(\sigma,\eta)}{\sigma}$ and $\hat{V}(\sigma,\eta) = \partial_\eta \hat{W} (\sigma,\eta) $. Consider the coordinates to range in
$0\leq \eta\leq P$, where $P$ is a real number, and $-\infty <\sigma <\infty$.
The differential equation (\ref{diffeq}) must be supplemented by boundary  and initial conditions. In terms of $\hat{W}(\sigma,\eta)$ the problem reads
\begin{eqnarray}
& & \partial^2_\sigma \hat{W}(\sigma,\eta)+\partial^2_\eta \hat{W}(\sigma,\eta)=0, \qquad \qquad \text{(almost everywhere)} \label{PDEhatv}\\
& &   \hat{W} (\sigma,\eta=0)=0,\;\;\;\;\;\hat{W} (\sigma,\eta=P)=0,\nonumber\\
& & \partial_\sigma \hat{W}(\sigma=0^+,\eta)- \partial_\sigma \hat{W}(\sigma=0^-,\eta)=- {\cal R}(\eta).\nonumber
\end{eqnarray}
As we discuss in the following sections, the function ${\cal R}(\eta)$ is the input  determined by the dual quiver field theory. Notice that, since $\hat{W}$ is an harmonic function, we have that also $\hat{V}$ is harmonic, which in turn implies \eqref{diffeq}.

To solve the problem in eq.(\ref{PDEhatv}), we separate variables and impose the boundary conditions to find,
\begin{equation}
\hat{V}(\sigma,\eta)= \sum_{k=1}^\infty a_k \cos\left( \frac{k\pi\eta}{P}\right) e^{-\frac{k \pi |\sigma|}{P}}, \qquad \hat{W}(\sigma,\eta)= \sum_{k=1}^\infty a_k \left(\frac{P}{k\pi}\right) \sin\left( \frac{k\pi\eta}{P}\right) e^{-\frac{k \pi |\sigma|}{P}}.\label{solutionPDE}
\end{equation}
We have used that the function ${\cal R}(\eta)$ has a Fourier decomposition,
\begin{equation}
{\cal R}(\eta)=\sum_{k=1}^\infty 2 a_k  \sin\left( \frac{k\pi\eta}{P}\right), \;\;\;\;\;a_k=\frac{1}{P}\int_0^P {\cal R}(\eta) \sin\left( \frac{k\pi\eta}{P}\right).\label{rankfunction}
\end{equation}
\subsection{Asymptotic behaviour}
Let us briefly study the asymptotic behaviour of our backgrounds. We start with the region $\sigma\to\pm\infty$.  
We combine the expressions in Appendix \ref{useful} together with the solutions in eq.(\ref{solutionPDE}). In particular, for $|\sigma|\to\infty$, we use eqs.(\ref{asymptsigma}) and write,
\begin{eqnarray}
& & ds^2_{st}\Big|_{\sigma\to\pm\infty}\sim \frac{\pi|\sigma|}{2} ds^2(\text{AdS}_4) +  \frac{d\eta^2}{2P } +\frac{P}{2 a_1}\sin^2\left(\frac{\pi \eta}{P}\right) d s^2 (S^2_1)+ \frac{\pi |\sigma|}{2} d s^2 (S^2_2)+ \frac{1}{2P}d\sigma^2,\nonumber\\
& & e^{-2\Phi}\sim \frac{e^{-\frac{2\pi |\sigma|}{P}} }{ |\sigma|}.\label{backgroundasympsigma}
\end{eqnarray}
Performing a change of coordinates $|\sigma| \to - \log r$, with $r$ small, one can notice that the metric and the dilaton highlights the presence of a $(p,q)$ five brane \cite{DHoker:2016ujz} with support on AdS$_4 \times S_2^2$. 

Let's now consider the asymptotic behaviour  at the physical boundary $\eta\sim 0,P$. We will explicit deal with the case $\eta \to 0$ since the discussion for the other boundary is identical. The expressions for $f_i(\sigma,0)$ can be read from eq.(\ref{asympeta}); schematically, we have
\begin{eqnarray}
& & ds^2_{st}\Big|_{(0,0)}\sim ds^2(\text{AdS}_4) +  d\eta^2 +\eta^2  d s^2 (S^2_1)+ d\sigma^2+ d s^2 (S^2_2) \, ,\;\;\;e^{-2\Phi}\sim 1 \, ,
\end{eqnarray}
where we have omitted functions of $\sigma$ which are not singular at least for $\sigma \neq 0$. Thus near to $\eta \sim 0$ we have a regular AdS$_4 \times S^2 \times \mathbb{R}^4 $ geometry. 

The behaviour in the corner $(\sigma,\eta)=(0,0)$ requires some more analysis, since in that case the functions in (\ref{asympeta}) can lead to a singular metric. Notice that, however, the dilaton is always finite, so if these singularities are present it would be hard to give them an interpretation in terms of branes. One possibility is to restrict our analysis to the $\mathcal{R}(\eta)$ which leads to a regular metric. We can do that by choosing $\mathcal{R}(\eta)$ to be linear near to $\eta = 0$, so that we have
\begin{equation}
\partial_\eta^2 \mathcal{R} |_{\eta \sim 0} = \partial_{\eta \sigma}^2 \hat{V} |_{(\sigma,\eta) \sim (0,0)} = 0 \qquad \Longleftrightarrow \qquad \sum_k k^3 a_k = 0 \, . \label{eq:Ransatz}
\end{equation}
With this ansatz, the metric has the following asymptotic behavior (see eq.\eqref{eq:asympetasigma} for the full expansion of the warping functions $f_i$)
\begin{eqnarray}
& & ds^2_{st}\Big|_{(0,0)}\sim ds^2(\text{AdS}_4) +  {d\eta^2} +\eta^2  d s^2 (S^2_1)+ d\sigma^2+ \sigma^2 d s^2 (S^2_2),\;\;\;e^{-2\Phi}\sim 1 ,\label{backgroundasympsigmaeta=0}
\end{eqnarray}
which is a regular AdS$_4 \times \mathbb{R}^6$ geometry.
It is particularly interesting to notice that requiring $\mathcal{R}(\eta)$ to be linear in $(\sigma,\eta)=(0,0)$ is not actually a restriction but a necessary condition for having the Page charges properly quantised, as we are going to see in the next section.

\section{Charges and other important quantities}\label{Pagechargessection}

%
To start, we write the expressions of the Page fluxes $\hat{F}_{p}= F_p\wedge e^{-B_2}$. For the Ramond and NS5 fields in our configuration of eq.(\ref{background}), we have
\begin{eqnarray}
& & \hat{F}_3=F_3, \qquad \hat{F}_5= F_5 - \left(B_2 -\frac{\pi \Delta}{2} \text{Vol}(S^2_1)\right)\wedge F_3.\label{pagefluxes}
\end{eqnarray}
Note that we have performed a large gauge transformation  $B_2\to\left( B_2 -\frac{\pi \Delta}{2} \text{Vol}(S^2_1) \right) $, that will be useful below.
The Page charges are defined as
\begin{equation}
Q_{Dp/NS5}=\frac{1}{(2\pi)^{7-p} \alpha'}\int_{\Sigma_{8-p} }\hat{F}_{8-p}.
\end{equation}
In the following, we will set $\alpha'=1$.
Let us study the charges associated with $H_3, \hat{F}_5, \hat{F}_3$.
\\

\underline{\bf NS5 branes charge}
\\
First, we analyse the charge of NS5 branes. We  choose a three-cycle to perform the integration of $H_3$,
\begin{equation}
\Sigma_3=[\eta, S_1^2(\theta_1,\varphi_1)]_{\sigma=\pm\infty}.
\end{equation}
We then find,
\begin{eqnarray}
& & Q_{NS5}=\frac{1}{4\pi^2 }\int_{\Sigma_3} H_3= \frac{1}{\pi}\int_{0}^{P}
\partial_\eta f_6(\sigma\to\pm\infty,\eta)= 
 P- \frac{\sigma \partial_\eta V \partial_\eta^2 V}{2\Lambda}\Big]_{ \sigma\to\pm\infty,0}^{\sigma\to\pm\infty, P}=P. \label{QNS5}
\end{eqnarray}
In the last two steps we have summed up the contributions at $\sigma = \pm \infty$ and used  eq.(\ref{forB2}).
In conclusion, the total number of NS5 branes is proportional to the length of the $\eta$-interval. 
\\

\underline{\bf D3 branes charge}
\\
To calculate the number of D3 branes we integrate  the expression 
for the Page flux $\hat{F}_5$ in eq.(\ref{pagefluxes}). 
The five manifold on which we integrate $\hat{F}_5$ is defined as
\begin{equation}
\Sigma_5=[S_1^2, S_2^2,\sigma]_{\eta= \text{fixed}}.
\end{equation}
We implemented a large gauge transformation as in \eqref{pagefluxes}, below we determine the parameter $\Delta$ to have a quantised number of D3 branes in each interval of the $\eta$-coordinate.

Using the potential $\hat{W}$ defined above eq.(\ref{PDEhatv}) we are able to compactly write the relevant component of $\hat{F}_5$ as,
\begin{eqnarray}
& & \hat{F}_5\Big|_{\Sigma_5}= F_5- (B_2- \frac{\pi}{2}\Delta \text{Vol}(S^2_1))\wedge F_3\Big|_{\Sigma_5}= \pi^2   \partial_\sigma \left({\cal M}_1+{\cal M}_2 \right)~\text{Vol}(S^2_1) \wedge \text{Vol}(S^2_1)\wedge d\sigma. \nonumber\\
& & {\cal M}_1=\frac{\eta\sigma (\partial_\eta^2 V) (\partial_\eta V) -\sigma (\partial_\eta V)^2}{\partial_{\eta\sigma}^2 V},\;\;\;\; {\cal M}_2= \partial_\sigma \left(\hat{W} - (\eta-\Delta) \partial_\eta \hat{W} \right). 
\end{eqnarray}
Then, we compute
\begin{equation}
N_{D3}=\frac{1}{(2\pi)^4 \alpha'}\int_{\Sigma_5} \hat{F}_5= 
\frac{ \pi^2\times (4\pi)^2}{(2\pi)^4 }\int d\sigma  \partial_\sigma\left( {\cal M}_1 +{\cal M}_2\right)=2\left( {\cal M}_1 +{\cal M}_2\right)\Big]_{\sigma=0^+,\eta}^{\sigma\to\infty,\eta}.
\end{equation}
Since $ {\cal M}_i$ are even, we have considered twice  the contribution for $\sigma >0$.

Using the expansions in eqs.(\ref{M1})-(\ref{M2}),
the reader can check that the contribution of ${\cal M}_1$ is vanishing at $\sigma\to\infty$ and at $\sigma=0$. 
Inspecting the expression for ${\cal M}_2$  in eq.(\ref{M2}) shows 
that  it vanishes for $\sigma\to\infty$.
In summary, the
charge of D3 branes is given by $\mathcal{M}_2$ evaluated at $\sigma = 0$, and using the definition of $\mathcal{R}$ in eq.\eqref{rankfunction} we get
\begin{equation}
N_{D3}= 2\partial_\sigma\left(\hat{W} - (\eta-\Delta) \partial_\eta \hat{W} \right)\Big|_{\sigma=0^+} = \mathcal{R}(\eta) - (\eta-\Delta) \mathcal{R}'(\eta)  .\label{charged3final}
\end{equation}
This expression indicates that the rank function used as input for the partial differential equation must be piecewise linear. In fact,  consider a  piecewise linear and continuous function defined in intervals
 \[ {\cal R}(\eta) = \begin{cases} 
          N_1 \eta & 0\leq \eta \leq 1 \\
          N_l+ (N_{l+1} - N_l)(\eta-l) & l \leq \eta\leq l+1,\;\;\; l:=1,...., P-2\\
  %
          N_{P-1}(P-\eta) & (P-1)\leq \eta\leq P .
       \end{cases}
    \]
The expression in eq.(\ref{charged3final}) indicates that, after choosing  $\Delta=k$ in the interval $[k, k+1]$ there are $N_k$ D3 branes.
\\

\underline{\bf  D5 brane charge}
\\
To calculate the charge of D5 branes, we use
\begin{equation}
N_{D5}=\frac{1}{4\pi^2 }\int_{\hat{\Sigma}_3} F_3,
\qquad
\hat{\Sigma}_3=[\eta, S_2^2(\theta_2,\varphi_2)]_{\sigma=0^+}.
\end{equation}
 We find,
 \begin{eqnarray}
 & & N_{D5}= -\frac{1}{\pi }\int_{\eta_i}^{\eta_f} d\eta \partial_\eta f_7(\sigma=0^+,\eta)=-\frac{1}{\pi} \left( f_7(0,\eta_f) - f_7(0,\eta_i)\right).\nonumber
 \end{eqnarray}
 Using eq.(\ref{f7}), the reader can check that 
  \begin{equation}
 N_{D5}=\mathcal{R}'(\eta_i)-\mathcal{R}'(\eta_f) .\label{D5charge}
 \end{equation}
 This result is expressing the number of D5s between the points $\eta_i$ and $\eta_f$ as computed by the differences in slope of the rank function at those two points. For a piece-wise continuous and linear  rank function as the one obtained in quantising the charge of D3 branes, we find that the charge of D5 branes is also quantised.
 
 In summary, the  rank-function  is the input for the PDE problem in eq.(\ref{PDEhatv}). To have quantised charges for Neveu-Schwarz five branes, we need the size of the interval $P$ to be an integer--consistently with the boundary conditions in eq.(\ref{PDEhatv}). To have quantised numbers of D3 and D5 branes, the rank function must be a piece-wise linear and continuous function of the form
  \[ {\cal R}(\eta) = \begin{cases} 
          N_1 \eta & 0\leq \eta \leq 1 \\
          N_l+ (N_{l+1} - N_l)(\eta-l) & l \leq \eta\leq l+1,\;\;\; l:=1,...., P-2\\
  %
          N_{P-1}(P-\eta) & (P-1)\leq \eta\leq P .
       \end{cases}
    \]
The number of D3   (colour) branes and D5 (flavour) branes in the interval $[k,k+1]$ and the total number of branes are given by,
\begin{eqnarray}
& & N_{D3}[k, k+1]= N_k,\;\;\;\; N_{D5}[k,k+1]=2 N_k - N_{k+1}- N_{k-1},\label{chargesk}\\
& & N_{D3}^\text{total}= \int_0^P {\cal R}(\eta) d\eta,\;\;\;\; N_{D5}^\text{total}= {\cal R}'(0) - {\cal R}'(P), \;\;\;\; N_{NS5}^\text{total}=P.\nonumber
\end{eqnarray}
\subsection{Hanany--Witten set-up and linking numbers}
The counting of branes described above encodes in the rank function ${\cal R}(\eta)$ the `kinematic data' of the dual conformal field theory. The presence of $P$ NS5 branes along the $\eta$-direction suggest that we should place one NS5 at each integer value of $\eta$. In between the $k^{th}$ and $(k+1)^{th}$ NS5-branes, we have $N_k$ D3 branes and $N_{F_k}=2 N_k - N_{k+1}- N_{k-1} $ D5 branes as indicated in eq.(\ref{chargesk}). Analysing the Ramond fields $\hat{F}_5$ and $\hat{F}_3$ suggests that the branes extend along the directions of space time as indicated in Table \ref{table1}.

\begin{figure}[t]
    \centering
    \begin{scriptsize}
   \begin{tikzpicture}[scale=.8]
   \node at (4,-2){(a)};
   
   \draw[thick](0,-.5)--(0,2.5);
   \node[label=below:{NS5$_1$}]at(0,-.5){};
   \node[label=above:{$F_2$ D5}][cross] at(3,1.8) {};
   \node[label=above:{$F_p$ D5}][cross] at(7,1.8) {};
   \node[label=above:{$F_1$ D5}][cross] at(1,1.8) {};
   \draw[thick](2,-.5)--(2,2.5);
   \node[label=below:{NS5$_2$}]at(2,-.5){};
   \node[label=below:{NS5$_3$}]at(4,-.5){};
   \node[label=below:{NS5$_{P-1}$}]at(6,-.5){};
   \node[label=below:{NS5$_P$}]at(8,-.5){};
   \node at (5,1.25) {$\cdots$};
   \draw[thick](4,-.5)--(4,2.5);
   \draw[thick](8,-.5)--(8,2.5);
   \draw[thick](6,-.5)--(6,2.5);
   \draw[thick](0,1)--(4,1);
   \node at (1,.6) {$N_1$ D3};
    \draw[thick](6,1)--(8,1);
    \node at (3,.6) {$N_2$ D3};
    \node at (7,.6) {$N_{p-1}$ D3};
   \end{tikzpicture}
   \end{scriptsize}
   \hspace{.25cm}
   \begin{scriptsize}
             \begin{tikzpicture}
                      \node[label=below:{$N_1$}][u](N1){};
                      \node[label=above:{$F_1$}][uf](F1)[above of=N1]{};
                      \node[label=below:{$N_2$}][u](N2)[right of=N1]{};
                      \node[label=above:{$F_2$}][uf](F2)[above of=N2]{};
                      \node (dots)[right of=N2]{$\cdots$};
                      \node[label=below:{$N_{P-1}$}][u](NP-1)[right of=dots]{};
                      \node[label=above:{$F_{P-1}$}][uf](FP-1)[above of=NP-1]{};
                      \draw(N1)--(F1);
                      \draw(N1)--(N2);
                      \draw(N2)--(F2);
                      \draw(N2)--(dots);
                      \draw(dots)--(NP-1);
                      \draw(NP-1)--(FP-1);
                      \node at (1.5,-2){(b)};
             \end{tikzpicture}
         \end{scriptsize}
  
    \caption{The Hanany--Witten brane set up, showing the $N_k$ (colour) D3 branes, the $F_k$ (flavour) D5 branes and the NS5 branes. The associated quiver field theory is also shown.}
    \label{quiver+HW}       
\end{figure}
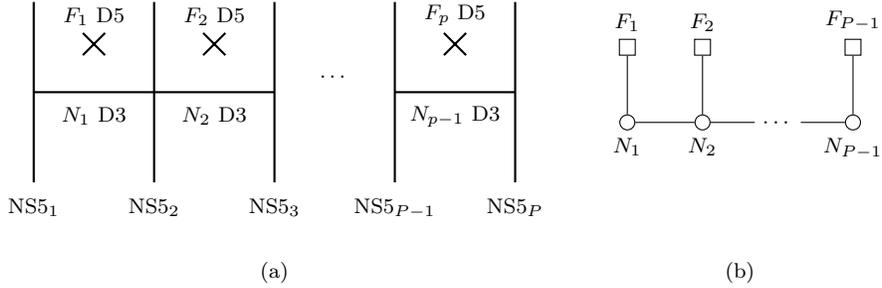

\begin{table}
    \centering
    \begin{tabular}{|c|c|c|c|c|c|c|c|c|c|c|}\hline 
         & $t$ &$x_1$ &$x_2$ &$r$ &$\theta_1$ &$\varphi_1$ &$\theta_2$&$\varphi_2$ &$\sigma$&$\eta$ \\\hline
         NS5 &$-$ & $-$& $-$& $\cdot$ & $\cdot$ & $\cdot$ &$-$ &$-$ &$-$ &$\cdot$\\\hline
         D5 &$-$ & $-$&$-$ &$-$ & $-$&$-$ & $\cdot$&$\cdot$ &$\cdot$ &$\cdot$\\\hline
         D3 &$-$ &$-$ &$-$ &$\cdot$ &$\cdot$ &$\cdot$ &$\cdot$ &$\cdot$ & $\cdot$&$-$\\\hline
    \end{tabular}
    \caption{The Hanany--Witten set up, indicating the directions over which each brane extends.}
    \label{table1}
\end{table}
In this Hanany--Witten set-up \cite{Hanany:1996ie},
the field theory is realised on the ${t, x_1,x_2}$ directions. The D3 branes have one compact direction leading to an effective $(2+1)$-dimensional dynamics, for each stack of $N_k$ branes, that give place to an $U(N_k)$ gauge group.
The D5 branes are effectively realising an $SU(N_{F_k})$ global symmetry, hence correspond to flavour branes. The NS5 branes provide the boundary conditions necessary for the D3 to end on them. We represent the system as in Figure \ref{quiver+HW}.
 One interesting quantity associated with these Hanany--Witten set-ups are the linking numbers. These are topological quantities (invariant under Hanany--Witten moves) associated with Neveu-Scharz and Ramond five branes. For the $i^{th}$ NS5 brane and the $j^{th}$ D5 brane they are defined in terms of the number of branes to the left and right of a given one,
 \begin{eqnarray}
 & & \hat{L}_{NS5i}= \left(n_{D3}^{\text{right}} - n_{D3}^{\text{left}}  \right) + n_{D5}^{\text{left}},\;\;\;\;\; {L}_{D5j}= \left(n_{D3}^{\text{right}} - n_{D3}^{\text{left}}  \right) + n_{NS5}^{\text{right}}.\label{linking1}
 \end{eqnarray}
For the systems described above, the linking numbers can be seen to have the values,
\begin{eqnarray}
& & \hat{L}_1=\hat{L}_2=....=\hat{L}_P= {\cal R}'(0), \;\;\;\;\; L_{D5,j}= P-j.\label{linking2}
\end{eqnarray}
These satisfy 
\begin{equation}
\sum_{i=1}^{N_{NS5}} \hat{L}_{i}= \sum_{j=1}^{N_{D5}} L_{D5,j}= \hat{N}.\label{linking3}
\end{equation}
Therefore with each quiver we associate two partitions of the integer $\hat{N}$. The partitions are made out of the linking numbers of NS5 and D5 branes,
\begin{equation}
\hat{\rho}=(\hat{L}_{NS1}, \hat{L}_{NS2},....,\hat{L}_{NSP}), \;\;\;\;\;\rho=(L_{D51}, L_{D52},...., L_{D5}).\label{linking4}
\end{equation}
The associated quiver field theories are referred to as $T_\rho^{\hat{\rho}} \left[ SU(\hat{N})\right]$. Gaiotto and Witten \cite{Gaiotto:2008ak} proposed
that these field theories flow to an interacting conformal point at low energies if a relation between partitions $\hat{\rho}^T\geq \rho$ is satisfied. The authors of the work \cite{Assel:2011xz} translated this condition into
$N_{D5,k}\geq 2 N_{D3,k} - N_{D3,k+1} - N_{D3,k-1}$. The formulation we presented in terms of a rank functions constrains us to the {\it balanced} quiver for which the equality is satisfied. We leave for future study the unbalanced situation.
\subsection{Holographic central charge}
Let us discuss now the holographic central charge. This is a quantity, instrumental in the tests of the duality between the backgrounds in (\ref{background}) and the conformal field theories described in the previous section.
The holographic central charge is defined as a weighted version of the volume of the  internal manifold (the part of the space that is not AdS$_4$). The definition of this quantity is carefully discussed in \cite{Macpherson:2014eza}-\cite{Bea:2015fja}, we refer the reader to those references for the general definitions. The application to our particular case is discussed below.

We use coordinates for AdS$_4$ such that
\begin{equation}
ds^2_{AdS_4}= e^{2\rho} dx_{1,2}^2+ d\rho^2.
\end{equation}
 We quote the relevant quantities that  can be read from the metric and dilaton in eq.(\ref{background})
\begin{eqnarray}
& &  a= f_1  (\sigma,\eta) e^{2\rho},\;\;\;\; b= e^{-2\rho},\;\;\;\; d=2,\nonumber\\
& &  ds_{int}^2=  f_1(\sigma,\eta)\Big[ f_2(\sigma,\eta) d s^2 (S^2_1)+ f_3(\sigma,\eta) d s^2 (S^2_2)+ f_4(\sigma,\eta)(d\sigma^2+d\eta^2) \Big],\nonumber\\
& &  \det[g_{int}]\!=\! f_1^6 f_2^2 f_3^2 f_4^2 \sin^2\theta_1 \sin^2\theta_2,\nonumber\\
& &  V_{int} = \int_{M_{int} } \sqrt{ \det{g_{int}} e^{-4 \Phi} a^d }  =  \left[ 16\pi^2 \int  d\sigma d\eta  f_1^4  f_2  f_3  f_4 f_5\right] e^{2\rho}={\cal N} e^{2\rho},\nonumber\\
& &  H= V_{int}^2={\cal N}^2 e^{4\rho}, \;\;\; H'= 4 H,\nonumber\\
& & c_{hol}=\frac{d^d}{G_N} b^{d/2} \frac{H^{\frac{2d+1}{2}} }{(H')^d}= \frac{{\cal N}}{4 G_N}= \frac{{\cal N}}{32\pi^6}.\label{centralformulas}
\end{eqnarray}
We have used $G_N=8\pi^6 \alpha'^4 g_s^2= 8\pi^6$ (in units where $\alpha'=g_s=1$). Using the definitions for the dilaton and the warp factors given in eq.(\ref{background}), we have
\begin{eqnarray}
& & {\cal N}=- 16 \pi^6 \int d\sigma d\eta (\sigma^2 \partial_\eta V)\partial_\sigma( \sigma \partial_\eta V)=-16 \pi^6 \int d\sigma d\eta ~\sigma (\partial_\eta \hat{V} )(\partial^2_{\sigma\eta}\hat{V}).
\end{eqnarray}
We use the expressions in Appendix \ref{useful}, perform explicitly the integral over $\eta$ and after that the $\sigma$-integral\footnote{Using that  
%
$\int_0^{P}  \sin\left( \frac{k\pi \eta}{P}\right)  \sin\left( \frac{l\pi \eta}{P}\right) d\eta= \frac{P}{2} \delta_{k,l}$.}.
We find,
\begin{equation}
{\cal N}= 4\pi^7 \sum_{k=1}^{\infty} k a_k^2,\;\;\;\;
c_{hol}= \frac{\pi}{8} \sum_{k=1}^\infty  k a_k^2 .\label{chol}
\end{equation}
Let us now evaluate explicitly this formula for a generic balanced quiver, characterised by a generic  rank function.
%
\subsubsection{Generic balanced quiver}\label{centralgenericbalanced}
In this section we derive an analytic expression for the holographic central charge in eq.(\ref{chol}) in the case of a generic quiver field theory. Consider a generic balanced 3d $\mathcal{N}=4$ linear quiver and its associated rank function
\begin{equation}\label{generic balanced quiver}
    \begin{array}{c}
         \begin{scriptsize}
             \begin{tikzpicture}
                      \node[label=below:{$N_1$}][u](N1){};
                      \node[label=above:{$F_1$}][uf](F1)[above of=N1]{};
                      \node[label=below:{$N_2$}][u](N2)[right of=N1]{};
                      \node[label=above:{$F_2$}][uf](F2)[above of=N2]{};
                      \node (dots)[right of=N2]{$\dotsb$};
                      \node[label=below:{$N_{P-1}$}][u](NP-1)[right of=dots]{};
                      \node[label=above:{$F_{P-1}$}][uf](FP-1)[above of=NP-1]{};
                      \draw(N1)--(F1);
                      \draw(N1)--(N2);
                      \draw(N2)--(F2);
                      \draw(N2)--(dots);
                      \draw(dots)--(NP-1);
                      \draw(NP-1)--(FP-1);
             \end{tikzpicture}
         \end{scriptsize}
    \end{array}\;;\quad R(\eta)=\left\{\begin{array}{cc}
         N_1\eta & \eta\in[0,1]  \\
        \vdots & \\
        N_k+(N_{k+1}-N_k)(\eta-k)& \eta\in[k,k+1]\\
        \vdots&\\
        N_{P-1}(P-\eta)&\eta\in[P-1,P]
    \end{array}\right.
\end{equation}
From the rank function we can compute the Fourier coefficients as defined in eq.(\ref{rankfunction})
\begin{eqnarray}
& &   a_k=\frac{1}{P}\sum_{j=0}^{P-1}\int_j^{j+1} \left[  N_j+(N_{j+1}-N_j)(\eta-j)   \right] \sin\left(\frac{k\pi \eta}{P}\right) d\eta\;, \quad \text{with}\; N_0=N_P=0 \, ,\nonumber\\
& & a_k=\frac{1}{\pi^2 k^2}\sum_{j=0}^{P-1}   k\pi \left[   N_j\cos\left(\frac{k\pi j}{P}\right)-N_{j+1}\cos\left(\frac{k\pi(j+1)}{P}\right)  \right]+ \nonumber\\
& &    P (N_{j+1}-N_j) \left[   \sin\left(\frac{k\pi(j+1)}{P}\right) -  \sin\left(\frac{k\pi j}{P}\right)  \right] \, .
 \label{genericak}
\end{eqnarray}
The first line of $a_k$ sums to zero. The second line, can be rewritten as
\begin{equation}
    a_k=\frac{P}{\pi^2 k^2}\sum_{j=0}^{P-1}F_j\sin\left(\frac{k\pi j}{P}\right)\;,
\end{equation}
where $F_j=2N_j-N_{j+1}-N_{j-1}$--here we used the {\it balanced} character of the quiver. Plugging this into \eqref{chol} we obtain our general formula for the holographic central charge to be
\begin{equation}\label{chol general balanced}
\begin{aligned}
 c_{hol}&=\frac{P^2}{8 \pi^3}\sum_{k=1}^\infty\sum_{j,l=0}^{P-1} \frac{F_j F_l}{k^3}\sin\left(\frac{k\pi j}{P}\right)\sin\left(\frac{k\pi l}{P}\right)\\
 &=-\frac{P^2}{32 \pi^3}\sum_{k=1}^\infty \sum_{j,l=0}^{P-1}\frac{F_jF_l}{k^3}\left(e^{\frac{i\pi k}{P}(j+l)}+e^{-\frac{i\pi k}{P}(j+l)}-e^{\frac{i\pi k}{P}(j-l)}-e^{-\frac{i\pi k}{P}(j-l)}\right)\\
 &=-\frac{P^2}{16 \pi^3}\sum_{j,l=0}^{P-1}  F_j F_l \text{Re}\left[ \text{Li}_3\left(e^{\frac{i\pi k}{P}(j+l)}\right)-
 \text{Li}_3\left(e^{\frac{i\pi k}{P}(j-l)}\right)\right] \, .
 \end{aligned}
\end{equation}
This expression should be compared with  equation (70) in the work  \cite{Coccia:2020wtk}, see also \cite{Raamsdonk:2020tin}. The authors  of  \cite{Coccia:2020wtk} derived a generic expression for the Free Energy on a three-sphere of a balanced quiver using localisation and matrix model methods. Just like it occurs in different dimensions, the holographic central charge is proportional to the Free Energy of the CFT on a sphere.

Had we considered a situation for which $N_0$,$N_P$ are in general nonzero, the expression in eq.(\ref{chol general balanced}) needs to be supplemented to include the effects of the two boundaries. We discuss this interesting situation  in Appendix \ref{chol-offset}.

In what follows, we will discuss some illustrative examples of balanced quivers. We will start giving the rank function, compute the Fourier coefficients, the  brane charges and the linking numbers of the brane system. We will precisely calculate the holographic central charge emphasising the scaling with the various parameters of the CFT. 
\section{Some examples}\label{examplessection}

The explicit discussion of examples gives the interested reader a better understanding of the formalism we developed. Also, it allows a more intuitive comprehension of the field theory kinematic and dynamical aspects. Below, we compute the various quantities for which we derived generic expressions in the previous sections. We discuss these quantities  in examples of increasing level of sophistication.

\subsection{Generic triangular rank function}\label{generictriangular}
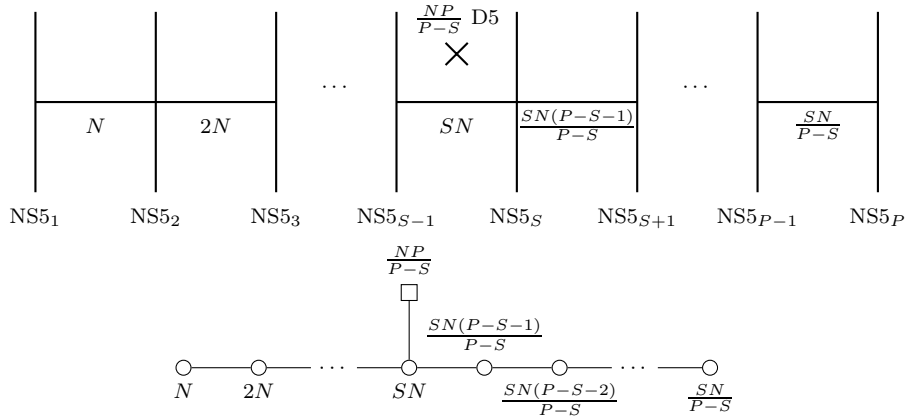
\begin{figure}[t]
    \centering
    \begin{scriptsize}
   \begin{tikzpicture}[scale=.8]
   
   \draw[thick](0,-.5)--(0,2.5);
   \node[label=below:{NS5$_1$}]at(0,-.5){};
   \node[label=above:{$\frac{NP}{P-S}$ D5}][cross] at(7,1.8) {};
   \draw[thick](2,-.5)--(2,2.5);
   \node[label=below:{NS5$_2$}]at(2,-.5){};
   \node[label=below:{NS5$_3$}]at(4,-.5){};
   \node[label=below:{NS5$_{S-1}$}]at(6,-.5){};
   \node[label=below:{NS5$_S$}]at(8,-.5){};
   \node[label=below:{NS5$_{S+1}$}]at(10,-.5){};
   \node[label=below:{NS5$_{P-1}$}]at(12,-.5){};
   \node[label=below:{NS5$_{P}$}]at(14,-.5){};
   \node at (5,1.25) {$\cdots$};
   \draw[thick](4,-.5)--(4,2.5);
   \draw[thick](8,-.5)--(8,2.5);
   \draw[thick](10,-.5)--(10,2.5);
   \draw[thick](12,-.5)--(12,2.5);
   \draw[thick](14,-.5)--(14,2.5);
   \draw[thick](6,-.5)--(6,2.5);
   \draw[thick](0,1)--(4,1);
   \node at (1,.6) {$N$};
    \draw[thick](6,1)--(10,1);
    \draw[thick](12,1)--(14,1);
    \node at (3,.6) {$2N$};
    \node at (7,.6) {$SN$};
    \node at (9,.6) {$\frac{SN(P-S-1)}{P-S}$};
    \node at (11,1.25) {$\cdots$};
    \node at (13,.6) {$\frac{SN}{P-S}$};
   \end{tikzpicture}
   \end{scriptsize}
   \hspace{.25cm}
   \begin{scriptsize}
             \begin{tikzpicture}
                      \node[label=below:{$N$}][u](N1){};
                      \node[label=below:{$2N$}][u](N2)[right of=N1]{};
                      \node (dots)[right of=N2]{$\cdots$};
                      \node[label=below:{$SN$}][u](NP-1)[right of=dots]{};
                      \node[label=above:{$\frac{SN(P-S-1)}{P-S}$}][u](P-S-1)[right of=NP-1]{};
                      \node[label=below:{$\frac{SN(P-S-2)}{P-S}$}][u](P-S-2)[right of=P-S-1]{};
                      \node (dots')[right of=P-S-2]{$\cdots$};
                      \node[label=below:{$\frac{SN}{P-S}$}][u](rightmost)[right of=dots']{};
                      \node[label=above:{$\frac{NP}{P-S}$}][uf](FP-1)[above of=NP-1]{};
                      \draw(N1)--(N2);
                      \draw(N2)--(dots);
                      \draw(dots)--(NP-1);
                      \draw(NP-1)--(FP-1);
                      \draw(NP-1)--(P-S-1);
                      \draw(P-S-1)--(P-S-2);
                      \draw(P-S-2)--(dots');
                      \draw(dots')--(rightmost);
             \end{tikzpicture}
         \end{scriptsize}
  
    \caption{The Hanany--Witten and quiver associated with a generic triangular rank function. As usual, vertical lines represent NS-five branes. Horizontal lines and circular nodes denote D3 branes and crosses and square nodes indicate D5 branes.}
    \label{quiver1+HW1}
\end{figure}

Our first example is described by the rank function
 \[ {\cal R}(\eta) = \begin{cases} 
          N \eta & 0\leq \eta \leq S \\
  %
          \frac{N S}{(P-S)}(P-\eta) & S \leq \eta\leq P ,
       \end{cases}
    \]
where we require $N / (P-S)$ to be integer, this condition will be needed to have properly quantised Page charges.
The first derivative of the rank function is
 \[ {\cal R}'(\eta) = \begin{cases} 
          N & 0\leq \eta \leq S \\
  %
          -\frac{N S}{(P-S)} & S \leq \eta\leq P ,
       \end{cases}
    \]
and  ${\cal R}''= \frac{N P}{(P-S)} \delta(\eta- S)$. The quiver and Hanany--Witten set-up associated with the rank function are given in Figure \ref{quiver1+HW1}.

The charge of D3 and D5 in each interval can be read from the rank function and its second derivative. The total number of branes follows from eq.(\ref{chargesk})
\begin{eqnarray}
& & Q_{D3}^\text{total}= \int_0^P {\cal R}(\eta) d\eta=\sum_{j=1}^{S} j N + \sum_{j=1}^{P-S}  \frac{N S}{(P-S) } (P-S-j)= \frac{N P S}{2} \, ,\label{chargesex1}\\
& & Q_{D5}^\text{total}= {\cal R}'(0)-{\cal R}'(P)= \frac{P N}{(P-S)},\;\;\;\;\; Q_{NS5}^\text{total}=P \, .\nonumber
\end{eqnarray}
For this family of quivers we calculate the Fourier coefficient of the rank function using eqs.(\ref{rankfunction}), (\ref{genericak}). We find,
\begin{equation}
a_k= \frac{N P^2}{(P-S) \pi^2 k^2}\sin\left( \frac{k\pi S}{P}\right).\label{akex1}
\end{equation}
The linking numbers can be computed using the definitions in eq.(\ref{linking1}), the Hanany--Witten set-up of Figure \ref{quiver1+HW1}, and the holographic expressions in eq.(\ref{linking2}),
\begin{eqnarray}
& & \hat{L}_{NS5_1}=\hat{L}_{NS5_2}=....=\hat{L}_{NS5_P}= {\cal R}'(0)= N,\label{linkingsex1}\\
& & L_{D5_1}=L_{D5_2}=....=L_{D5_{PN/(P-S)}}= P-i= P-S.\nonumber
\end{eqnarray}
These values satisfy the relation in eq.(\ref{linking3}), $\sum_{NS5} \hat{L}_i=\sum_{D5} L_{j}= N P$. These numbers define two partitions of $\hat{N}= N P$,
\begin{equation}
\hat{\rho}=(N,N,N,N....,N)=([N]^P),\;\;\;\; \rho=(P-S, P-S,...., P-S)=\left([P-S]^{\frac{PN}{(P-S)}}\right),
\end{equation}
and the quiver in Figure \ref{quiver1+HW1} represents the theory $T_{\rho}^{\hat{\rho} } [SU(N P)]$.

We can compute the holographic central charge using eq.(\ref{chol}) and the Fourier coefficient in eq.(\ref{akex1}),
\begin{equation}
c_{hol}= \frac{N^2 P^4}{32\pi^3 (P-S)^2} \left[     2\zeta(3)- 2 \text{Re}~ \text{Li}_3(e^{\frac{2\pi i S}{P} } )        \right].\label{cholex1}
\end{equation}
This  family of quivers have some interesting special cases. Indeed, consider the case $S=(P-1)$, the expressions derived in eqs.(\ref{chargesex1})-(\ref{cholex1}) are valid. Interestingly, in the holographic limit ($P$ being very large), we find
\begin{equation}
\lim_{P\to\infty}c_{hol}= \frac{N^2 P^2}{8\pi}\log P.
\end{equation}
Another interesting case is the `symmetric quiver' for which $2 S=P$. In this case we find,
\begin{equation}
\lim_{P\to\infty}c_{hol}=\frac{7 N^2 P^2}{16\pi^3}\zeta(3).\label{simxx}
\end{equation}
It is also interesting the case in which $S$ is some fixed integer, not scaling with $P$. The holographic limit for this situation gives,
\begin{equation}
\lim_{P\to\infty}c_{hol}=\frac{N^2 S^2}{8\pi}\log\left(P \right).
\end{equation}
The expression of the holographic central charge and its limiting cases clearly display the non-perturbative character of the result. Let us analyse a more elaborated example.

\subsection{Generic trapezoidal rank function}\label{trapezoidalgeneric}
\begin{figure}[t]
    \centering
    \begin{scriptsize}
   \begin{tikzpicture}[scale=.7]
   
   \draw[thick](0,-.5)--(0,2.5);
   \node[label=below:{$1$}]at(0,-.5){};
   \node[label=above:{$N$ D5}][cross] at(7,1.8) {};
   \node[label=above:{$\frac{MN}{Q}$ D5}][cross] at(13,1.8) {};
   \draw[thick](2,-.5)--(2,2.5);
   \node[label=below:{$2$}]at(2,-.5){};
   \node[label=below:{$3$}]at(4,-.5){};
   \node[label=below:{${M-1}$}]at(6,-.5){};
   \node[label=below:{$M$}]at(8,-.5){};
   \node[label=below:{${M+1}$}]at(10,-.5){};
   \node[label=below:{${M+S-1}$}]at(12,-.5){};
   \node[label=below:{${M+S}$}]at(14,-.5){};
   \node[label=below:{${M+S+1}$}]at(16,-.5){};
   \node[label=above:{${M+S+Q-1}$}]at(18,2.5){};
   \node[label=below:{${M+S+Q}$}]at(20,-.5){};
   \node at (5,1.25) {$\cdots$};
   \draw[thick](4,-.5)--(4,2.5);
   \draw[thick](8,-.5)--(8,2.5);
   \draw[thick](10,-.5)--(10,2.5);
   \draw[thick](12,-.5)--(12,2.5);
   \draw[thick](14,-.5)--(14,2.5);
   \draw[thick](16,-.5)--(16,2.5);
   \draw[thick](18,-.5)--(18,2.5);
   \draw[thick](20,-.5)--(20,2.5);
   \draw[thick](6,-.5)--(6,2.5);
   \draw[thick](0,1)--(4,1);
   \node at (1,.6) {$N$};
    \draw[thick](6,1)--(10,1);
    \draw[thick](12,1)--(16,1);
    \draw[thick](20,1)--(18,1);
    \node at (3,.6) {$2N$};
    \node at (7,.6) {$MN$};
    \node at (9,.6) {$MN$};
    \node at (11,1.25) {$\cdots$};
    \node at (13,.6) {$MN$};
    \node at (15,.6) {$\frac{MN(Q-1)}{Q}$};
    \node at (17,1.25) {$\cdots$};
    \node at (19,.6) {$\frac{MN}{Q}$};
   \end{tikzpicture}
   \end{scriptsize}
   \hspace{.25cm}
   \begin{scriptsize}
             \begin{tikzpicture}
                      \node[label=below:{$N$}][u](N1){};
                      \node[label=below:{$2N$}][u](N2)[right of=N1]{};
                      \node (dots)[right of=N2]{$\cdots$};
                      \node[label=below:{$MN$}][u](MN)[right of=dots]{};
                      \node[label=below:{$MN$}][u](MN')[right of=MN]{};
                      \node (dots')[right of=MN']{$\cdots$};
                      \node[label=below:{$MN$}][u](MN'')[right of=dots']{};
                      \node[label=above:{$\frac{MN(Q-1)}{Q}$}][u](MNQ1)[right of=MN'']{};
                      \node[label=below:{$\frac{MN(Q-2)}{Q}$}][u](MNQ2)[right of=MNQ1]{};
                      \node (dots'')[right of=MNQ2]{$\cdots$};
                      \node[label=below:{$\frac{MN}{Q}$}][u](MNQQ)[right of=dots'']{};
                      \node[label=above:{$N$}][uf](F1)[above of=MN]{};
                      \node[label=above:{$\frac{MN}{Q}$}][uf](F2)[above of=MN'']{};
                      \draw(N1)--(N2);
                      \draw(N2)--(dots);
                      \draw(dots)--(MN);
                      \draw(MN)--(MN');
                      \draw(MN')--(dots');
                      \draw(dots')--(MN'');
                      \draw(MN'')--(MNQ1);
                      \draw(MNQ1)--(MNQ2);
                      \draw(MNQ2)--(dots'');
                      \draw(dots'')--(MNQQ);
                      \draw(MN)--(F1);
                      \draw(MN'')--(F2);
                      \draw [thick,decorate,decoration={brace,amplitude=6pt,mirror,raise=20 pt},xshift=0pt,yshift=10pt]
(MN) -- (MN'')node [black,midway,xshift=0pt,yshift=-35pt] {
$S$ nodes};
             \end{tikzpicture}
         \end{scriptsize}
  
    \caption{The Hanany--Witten set-up and the associated quiver for the trapezoidal rank function. The conventions are those described previously.}
    \label{quiver2+HW2}
\end{figure}
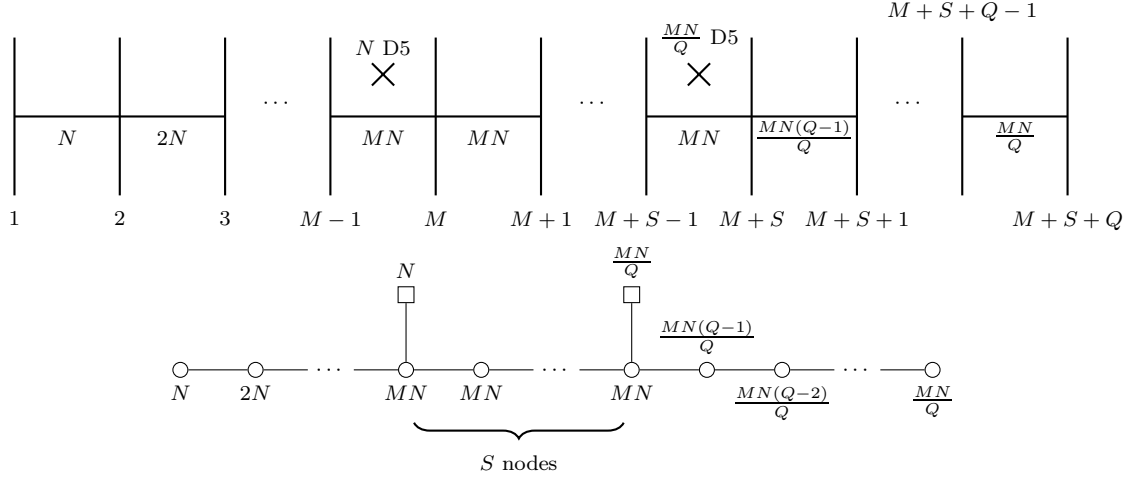

The rank function corresponding to this more sophisticated example is,
 \[ {\cal R}(\eta) = \begin{cases} 
          N \eta & 0\leq \eta \leq M \\
  %
         N M & M \leq \eta\leq M+S\\
          \frac{M N}{Q}(M+S+Q-\eta) & M+S \leq \eta\leq M+S+Q .
       \end{cases}
    \]
In this case $P=M+S+Q$ and we also require that $MN$ is a even multiple of $Q$.
The second derivative of the rank function is

\begin{equation}
{\cal R}''= N \delta(\eta-M) + \frac{M N}{Q} \delta(\eta- S-M).
\end{equation}
 The quiver and Hanany--Witten set-up associated with the rank function are given in Figure \ref{quiver2+HW2}.
The reader can check that for both examples the balanced-quiver condition is satisfied. 
Let us perform the same calculations we did in the previous example.

The charges of D3 and D5 in each interval can be read from the rank function and its second derivative. The total number of branes follow from eq.(\ref{chargesk})
\begin{eqnarray}
& & Q_{D3}^\text{total}= \int_0^P {\cal R}(\eta) d\eta=\sum_{j=1}^{M} j N + N M S+ \sum_{j=1}^{Q-1}  \frac{N M}{Q } (Q-j)= \frac{N M}{2}(P+S)\nonumber\\
& & Q_{D5}^\text{total}= {\cal R}'(0)-{\cal R}'(P)=N+ \frac{M N}{Q},\;\;\;\;\; Q_{NS5_5}^\text{total}=P=M+S+Q. \label{chargesex2}
\end{eqnarray}
We calculate the Fourier coefficient of the rank function using eq.(\ref{genericak}). We find,
\begin{equation}
a_k= \frac{N P}{Q \pi^2 k^2}\left[ Q \sin\left( \frac{k\pi M}{P}\right) + M\sin\left( \frac{k\pi(M+S)}{P}\right) \right].\label{akex2}
\end{equation}
The linking numbers can be computed using the definitions in eq.(\ref{linking1}),  the Hanany--Witten set-up of Figure \ref{quiver2+HW2}, and the holographic expressions in eq.(\ref{linking2}),
\begin{eqnarray}
& & \hat{L}_{NS5_1}=\hat{L}_{NS5_2}=....=\hat{L}_{NS5_P}= {\cal R}'(0)= N,\label{linkingsex2}\\
& & L_{D5_1}=L_{D5_2}=....L_{D5_{N}}=P-i= (S+Q),\nonumber\\
& &  L_{D51'}=L_{D52'}=....=L_{D5'_{MN/Q}}= P-i= Q.\nonumber
\end{eqnarray}
We have two stacks of D5 branes (distinguished by a $'$-symbol). These are located at $i=M$ and $i'= M+S$. These values for the linking numbers satisfy the relation in eq.(\ref{linking3}), $\sum_{NS5} \hat{L}_i=\sum_{D5} L_{j}= N (M+S+Q)= P N$. These numbers define two partitions of $\hat{N}= N P$,
\begin{eqnarray}
& & \hat{\rho}=(N,N,N,N....,N)=\left([N]^P \right),\nonumber\\
&& \rho=(S+Q, S+Q,...., S+Q; Q,Q,....Q)=\left([S+Q]^N; [Q]^{\frac{M N}{Q}}\right),\nonumber
\end{eqnarray}
and the quiver in Figure \ref{quiver2+HW2} represents the theory $T_{\rho}^{\hat{\rho} } [SU(N (M+S+Q))]$.

We can compute the holographic central charge using eq.(\ref{chol}) and the Fourier coefficient in eq.(\ref{akex2}),
\begin{eqnarray}
& & c_{hol}= \frac{N^2 (M+Q+S)^2}{16\p^3 Q^2} \text{Re}\Big[     (M^2+Q^2)\zeta(3) -Q^2 ~ \text{Li}_3(e^{\frac{2\pi i M}{P} } ) - M^2~ \text{Li}_3(e^{\frac{2 i \pi (M+S)}{P}})\nonumber\\
& & -2 M Q ~\text{Li}_3(e^{\frac{i \pi(2M+S)}{P}}) +2MQ \text{Li}_3(e^{\frac{i \pi S}{P}})        \Big].\label{cholex2}
\end{eqnarray}
 The holographic limit ($P=M+Q+S$ being very large) is more subtle than in the previous example as we can take $M$ very large, keeping fixed  $Q,S$ and the other two combinations. We find
\begin{eqnarray}
& &\lim_{M\to\infty}c_{hol}= \frac{N^2 M^2}{8\pi}\log \left( M \right) ,\;\;\;\; Q, S ~\text{are fixed}.\nonumber\\
& & \lim_{S \to\infty}c_{hol}= \frac{N^2 M^2}{8\pi}\log\left(S^2\right) ,\;\;\;\; Q, M ~\text{are fixed}.\nonumber\\
& & \lim_{Q\to\infty}c_{hol}= \frac{N^2 M^2}{8\pi}\log\left( Q \right) ,\;\;\;\; M, S ~\text{are fixed}.\nonumber
\end{eqnarray}
Another interesting situation is the `symmetric quiver' for which $Q=M$ and $P=2Q+S$. In this case we find,
\begin{eqnarray}
& & c_{hol}=\frac{N^2 (2Q+S)^2}{32\pi^3 } \text{Re}\Big[     7\zeta(3) - 2 \text{Li}_3(e^{\frac{2\pi i Q}{P} } ) - 2 \text{Li}_3(e^{\frac{2 i \pi (Q+S)}{P}}) + 4  \text{Li}_3(e^{\frac{i \pi S }{P} })       \Big].\nonumber\\
& &\lim_{Q\to\infty}c_{hol}= \frac{7}{4\pi^3}Q^2 N^2\zeta(3) ,\;\;\;\; S ~\text{is fixed}.\nonumber\\
& & \lim_{S \to\infty}c_{hol}= \frac{N^2 Q^2}{4\pi} \log\left( S \right) ,\;\;\;\; Q~\text{is fixed}.\label{limitssymmetric}
\end{eqnarray}
As a consistency check, notice that the second result in eq.(\ref{limitssymmetric}) is the same as that in eq.(\ref{simxx}) for $Q=\frac{P}{2}$ and $S=0$.

\section{Mirror Symmetry}\label{sectionmirrorxx}
 Many 3d $\mathcal{N}=4$ gauge theories enjoy a duality known as mirror symmetry \cite{Intriligator:1996ex,Kapustin:1999ha}. Our goal in this section is to provide a holographic perspective on mirror symmetry through some of the machinery introduced in the previous sections. In order to keep the discussion self-contained, we will proceed by recalling relevant aspects of 3d $\mathcal{N}=4$ supersymmetry and introduce the notion of mirror symmetry. We will then go over how mirror symmetry is derived from the Hanany--Witten setup by studying a specific example and provide consistency checks from holography, by matching the holographic central charges of the mirror pair.

The 3d $\mathcal{N}=4$ supersymmetry algebra admits two short representations known as vector multiplets and hypermultiplets respectively. The bosonic components of a vector multiplet include a gauge field and 3 real scalars, while the bosonic fields in a hypermultiplet are comprised of two complex (or four real) scalars. Under the SU$(2)_C\times$SU$(2)_H$ R-symmetry, the scalars in the vector multiplet form a $(\textbf{3},\textbf{1})$, while the hypermultiplet scalars form a $(\textbf{1},\textbf{2})$.  A 3d $\mathcal{N}=4$ gauge theory is specified by a choice of gauge group $G$, to which one associates a vector multiplet in the adjoint representation of $G$, as well as choice of matter content, specified by hypermultiplets transforming in representation $\rho$ of $G$. Since Maxwell's theory in 3d is dual to a periodic scalar, one can trade out the gauge field component of a free vector multiplet with another scalar field. Upon doing so, the field content of a vector multiplet and hypermultiplet now become almost indistinguishable.\footnote{One might be worried about the fact that the dual scalar is a compact scalar. Indeed the dual photon is an $S^1$-valued scalar, where the radius of the $S^1$ is proportional to the gauge coupling $g^2$. However, in the infra-red limit $g^2\rightarrow\infty$ this scalar decompactifies and one has 4 real-valued scalars in the vectormultiplet.} The only way one can tell them apart is by their transformation under the R-symmetry group, and the dualised vector multiplet is referred to as a twisted hypermultiplet. This may be viewed as a precursor, or hint of mirror symmetry; mirror symmetry is a non-trivial generalisation of this curious observation about free vector multiplets and hypermultiplets, but now applied to interacting quantum field theories. The moduli space of vacua of a 3d $\mathcal{N}=4$ theory is generically comprised of a Coulomb branch $\mathcal{M}_C$, a Higgs branch $\mathcal{M}_H$ and a mixed branch $\mathcal{M}_{\text{mix}}$. The Higgs branch, parameterised by VEVs of scalars in the hypermultiplet, is protected by a holomorphic non-renormalisation theorem \cite{Argyres:1996eh}, and as such is classically exact. The Coulomb branch is classically parameterised by VEVs of scalars in the twisted hypermultiplet, which are the coordinates of the Coulomb branch at large VEVs. However in the quantum theory one has to replace the complex scalar built out of one of the 3 scalars and the dual photon by a BPS monopole operator. Denoting by $n_h$ and $n_v$ the number of hypermultiplets and vector multiplets respectively, the quaternionic dimension of the Higgs and Coulomb branches for a generic linear quiver \eqref{generic balanced quiver} are given by
\begin{equation}
\begin{aligned}
\dim \left(\mathcal{M}_C\right)&=n_v=\sum_{i=1}^{P-1}N_i^2\;;\\
    \dim \left(\mathcal{M}_H\right)&=n_h-n_v=\sum_{i=1}^{P-1}N_iF_i+\sum_{i=1}^{P-2}N_iN_{i+1}-\sum_{i=1}^{P-1}N_i^2\;,
    \end{aligned}
\end{equation}
where in the second line, the first sum is the contribution of fundamental hypers attached to each node, whereas the second sum counts bi-fundamental hypers between neighbouring gauge nodes.
Gauge theories in 3d enjoy a topological or magnetic symmetry associated with the current 
\begin{equation}
    J_\text{top}=\star\tr{F}\;,
\end{equation}
whose conservation follows from the Bianchi identity. For quiver theories of the type we are interested in \eqref{generic balanced quiver}, there is one such conserved current for each gauge group factor and so the magnetic symmetry is classically $G_C^\text{cl}=$U$(1)^{P-1}$, while the flavour symmetry is given by $G_H=\text{S}\left[\prod_{i=1}^{P-1}\text{U}\left(F_i\right)\right]$, together with the R-symmetry the full classical 0-form global symmetry of quiver theories of the type \eqref{generic balanced quiver} is
\begin{equation}
    \text{U}(1)^{P-1}\times\text{S}\left[\prod_{i=1}^{P-1}\text{U}\left(F_i\right)\right]\times\text{SU}(2)_C\times\text{SU}(2)_H\;.
\end{equation}
In the quantum theory, the magnetic symmetry can be enhanced to a non-abelian symmetry. In \cite{Gaiotto:2008ak} Gaiotto and Witten conjectured the pattern of enhancement of the magnetic symmetry by analysing the monopole spectrum. Their conjecture states that for a quiver of the type  in eq.\eqref{generic balanced quiver}, whenever a chain of $n_i$ adjacent nodes are balanced, there is an enhancement of the form $\text{U}(1)^{n_I}\subset\text{SU}(n_I+1)$. Hence the full quantum 0-form symmetry of such quivers takes the form
\begin{equation}
    \text{U}(1)^{P-1-\sum_{I\in B}n_I}\times\prod_{I\in B}\text{SU}(n_I+1)\times\text{S}\left[\prod_{i=1}^{P-1}\text{U}\left(F_i\right)\right]\times\text{SU}(2)_C\times\text{SU}(2)_H\;,
\end{equation}
where the index $I$ takes values in the set $B$ of chains of balanced nodes. 
Mirror symmetry then relates pairs of 3d $\mathcal{N}=4$ theories where the flavour symmetry on one side is manifest as the magnetic or topological symmetry on the magnetic side. Moreover the two R-symmetry factors as well as the Coulomb and Higgs branches are also exchanged under mirror symmetry. 

From the type IIB perspective, mirror symmetry is a consequence of S-duality \cite{Hanany:1996ie}. The flavour symmetry of the low energy 3d theory is given by the gauge symmetry on the D5s, while its topological symmetry in the string embedding corresponds to the gauge symmetry on the NS5 worldvolume. Indeed, this is consistent with the field theory expectation since S-duality exchanges D5s and NS5s, while leaving D3 branes invariant. Note that in the type IIB embedding the R-symmetry is realised as the SO$(3)_C\times$SO$(3)_H$ rotational symmetry of the three coordinates along the NS5 (respectively D5) worldvolume which are transverse to the worldvolume directions of the D3 brane. Holographically this corresponds to the isometry of the two $S^2$ factors in the background. Hence we see that in order to realise mirror symmetry, one needs to accompany S-duality with a spacetime rotation.

Let us see how one can derive the mirror of the balanced quiver with a generic triangular rank function from its Hanany--Witten configuration in Figure \ref{quiver1+HW1}. In order to read off the mirror, we first perform a series of Hanany--Witten transitions on the brane system in Figure \ref{quiver1+HW1}. In particular, we first move all D5 branes to the right of all the NS5 branes, keeping in mind that whenever a D5 crosses an NS5, a D3 brane suspended between them is created
\begin{equation}
\begin{array}{c}
    \begin{scriptsize}
   \begin{tikzpicture}[scale=.8]
   
   \draw[thick](0,-.5)--(0,2.5);
   \node[label=below:{NS5$_1$}]at(0,-.5){};
   \draw[thick](2,-.5)--(2,2.5);
   \node[label=below:{NS5$_2$}]at(2,-.5){};
   \node[label=below:{NS5$_3$}]at(4,-.5){};
   \node[label=below:{NS5$_{P-1}$}]at(6,-.5){};
   \node[label=below:{NS5$_P$}]at(8,-.5){};
   
   \node at (5,1.25) {$\cdots$};
   \draw[thick](4,-.5)--(4,2.5);
   \draw[thick](8,-.5)--(8,2.5);
   \draw[thick](6,-.5)--(6,2.5);
   \draw[thick](0,1)--(4,1);
   \node at (1,.6) {$N$};
    \draw[thick](6,1)--(12,1);
    \draw[thick](18,1)--(14,1);
    \node[cross]at(10,1){};
    \node[cross]at(12,1){};
    \node[cross]at(14,1){};
    \node[cross]at(16,1){};
    \node[cross]at(18,1){};
    \node at (3,.6) {$2N$};
    \node at (7,.6) {$(P-1)N$};
    \node at (9,.6) {$PN$};
    \node at (11,.6) {$P(N-1)+S$};
    \node at (15,.6) {$2(P-S)$};
    \node at (17,.6) {$(P-S)$};
    \node at (13,1) {$\cdots$};
   \end{tikzpicture}
   \end{scriptsize}
\end{array}\;.
\end{equation}
Next, we perform S-duality, together with a spacetime rotation which exchanges D5s and NS5s from the above configuration to arrive at
\begin{equation}
\begin{array}{c}
    \begin{scriptsize}
   \begin{tikzpicture}[scale=.8]
   
   \node[cross] at(0,1){};
   \node[label=below:{NS5$_1$}]at(18,-.5){};
   \node[cross] at(2,1){};
   \node[label=below:{NS5$_2$}]at(16,-.5){};
   \node[label=below:{NS5$_3$}]at(14,-.5){};
   \node[label=below:{NS5$_{\frac{NP}{P-S}-1}$}]at(12,-.5){};
   \node[label=below:{NS5$_\frac{NP}{P-S}$}]at(9.5,-.5){};
   
   \node at (5,1) {$\cdots$};
   \node[cross] at(4,1){};
   \node[cross] at(8,1){};
   \node[cross] at(6,1){};
   \draw[thick](0,1)--(4,1);
   \node at (1,.6) {$N$};
    \draw[thick](6,1)--(12,1);
    \draw[thick](18,1)--(14,1);
    \draw[thick](9.5,-.5)--(9.5,2.5){};
    \draw[thick](12,-.5)--(12,2.5){};
    \draw[thick](14,-.5)--(14,2.5){};
    \draw[thick](16,-.5)--(16,2.5){};
    \draw[thick](18,-.5)--(18,2.5){};
    \node at (3,.6) {$2N$};
    \node at (7,.6) {$(P-1)N$};
    \node at (9,.6) {$PN$};
    \node at (10.75,.6) {$P(N-1)+S$};
    \node at (15,.6) {$2(P-S)$};
    \node at (17,.6) {$(P-S)$};
    \node at (13,1) {$\cdots$};
   \end{tikzpicture}
   \end{scriptsize}
\end{array}\;.
\end{equation}
In order to read off a gauge theory from this mirror Hanany--Witten setup, we recall the s-rule, which states that there can be at most 1 D3 brane suspended between a D5 and an NS5 brane, or else supersymmetry is broken. Since each of the D5 branes in the above configurations has $N$ D3 branes ending on it, the only way to satisfy the s-rule would be to have one of the $N$ D3s ending on NS5$_\frac{NP}{P-S}$, one D3 ending on NS5$_{\frac{NP}{P-S}-1}$ and so on, such that the last D3 brane ends on NS5$_{\frac{NP}{P-S}-N+1}$. We can move all $P$ D5 branes to the segment in between NS5$_{\frac{NP}{P-S}-N+1}$ and NS5$_{\frac{NP}{P-S}-N}$, keeping in mind that in doing so, one annihilates all D3 branes suspended between the D5 branes and the NS5 branes. The resulting configuration is     
\begin{equation}
    \begin{array}{c}
         \begin{scriptsize}
             \begin{tikzpicture}[scale=.8]
                      \draw[thick](0,-.5)--(0,2.5);
                      \draw[thick](2,-.5)--(2,2.5);
                      \draw[thick](4,-.5)--(4,2.5);
                      \draw[thick](6,-.5)--(6,2.5);
                      \draw[thick](8,-.5)--(8,2.5);
                      \draw[thick](10,-.5)--(10,2.5);
                      \draw[thick](12,-.5)--(12,2.5);
                      \draw[thick](14,-.5)--(14,2.5);
                      \draw[thick](0,1)--(4,1);
                      \node at (5,1){$\cdots$};
                      \draw[thick](6,1)--(8,1);
                      \node at (9,1){$\cdots$};
                      \draw[thick](10,1)--(14,1);
                      \node[label=above:{NS5$_1$}]at(14,2.5){};
                      \node[label=below:{NS5$_2$}]at(12,-.5){};
                      \node[label=above:{NS5$_3$}]at(10,2.5){};
                      \node[label=below:{NS5$_{\frac{NP}{P-S}-N}$}]at(8,-.5){};
                      \node[label=above:{NS5$_{\frac{NP}{P-S}-N+1}$}]at(6,2.5){};
                      \node[label=below:{NS5$_{\frac{NP}{P-S}-2}$}]at(4,-.5){};
                      \node[label=above:{NS5$_{\frac{NP}{P-S}-1}$}]at(2,2.5){};
                      \node[label=below:{NS5$_{\frac{NP}{P-S}}$}]at(0,-.5){};
                      \node[label=below:{$S$}]at(1,1){};
                      \node[label=below:{$2S$}]at(3,1){};
                      \node[label=below:{$NS$}]at(7,1){};
                      \node[label=below:{$2(P-S)$}]at(11,1){};
                      \node[label=below:{$(P-S)$}]at(13,1){};
                      \node[label=above:{$P$}][cross] at (7,1.75){};
             \end{tikzpicture}
         \end{scriptsize}
    \end{array}\;.
\end{equation}
From here one can immediately read off the mirror quiver
\begin{equation}
    \begin{array}{c}
         \begin{scriptsize}
             \begin{tikzpicture}
                      \node[label=below:{$S$}][u](S){};
                      \node[label=below:{$2S$}][u](2S)[right of=S]{};
                      \node(dots)[right of=2S]{$\cdots$};
                      \node[label=below:{$NS$}][u](NS)[right of=dots]{};
                      \node(dots')[right of=NS]{$\cdots$};
                      \node[label=above:{$2(P-S)$}][u](2P-2S)[right of=dots']{};
                      \node[label=below:{$(P-S)$}][u](P-S)[right of=2P-2S]{};
                      \node[label=above:{$P$}][uf](F)[above of=NS]{};
                      \draw(S)--(2S);
                      \draw(2S)--(dots);
                      \draw(dots)--(NS);
                      \draw(NS)--(dots');
                      \draw(NS)--(F);
                      \draw(dots')--(2P-2S);
                      \draw(2P-2S)--(P-S);
             \end{tikzpicture}
         \end{scriptsize}
    \end{array}\;.
\end{equation}
Note that this quiver has an SU$(P)$ flavour symmetry, while its magnetic symmetry is SU$\left(\frac{NP}{P-S}\right)$, due to the fact that there are $\frac{NP}{P-S}-1$ balanced gauge nodes.
On the other hand, the quiver in Figure \ref{quiver1+HW1} has an SU$\left(\frac{NP}{P-S}\right)$ flavour symmetry, while its magnetic symmetry is SU$(P)$ since there are $P-1$ balanced gauge nodes. We see that the flavour and magnetic symmetries are exchanged under the mirror map, as expected.

It would be natural to explore mirror symmetry for more involved quivers, for instance those with more than a single flavour node. However with a little thought, one immediately comes across the following observation. 
\begin{lemma}
The mirror of a quiver with more than one flavour node is necessarily unbalanced.
\end{lemma}
This is an immediate corollary of Gaiotto and Witten's global symmetry conjecture reviewed above. Recall that, under the mirror map, flavour symmetry is mapped to the magnetic symmetry of the dual theory. Thus the flavour symmetry of the electric theory, which is a product due to the fact that there are multiple flavour nodes, must be realised as the magnetic symmetry of the magnetic theory. The magnetic symmetry can only take a product form if there are distinct sets of balanced chains of gauge nodes. Under the assumption that the mirror theory is not a product of decoupled quivers,\footnote{That this can be a possibility is motivated by \cite{Akhond:2021knl}, where several families of quiver gauge theories were studied and found to have a factorised structure on their moduli space.} the only way to have such a situation is if there is at least one unbalanced node in between any two distinct chains of balanced nodes.
\subsection{Geometry and Mirror Symmetry}
 
 In other formulations of holographic duals to ${\cal N}=4$ three dimensional SCFTs, mirror symmetry manifests itself as S-duality.
 This is the case for the formulation of  \cite{DHoker:2007hhe}, \cite{Assel:2011xz}, based on two holomorphic functions
${\cal A}_1,{\cal A}_2$.  We choose ${\cal A}_2\sim z$, as in eq.(\ref{B7}) and in this choice that the connection between S-duality and mirror symmetry fades away in 
our formulation, indeed one can check that by choosing $\mathcal{A}_1 \sim z$ one obtains the background S-dual to $\mathcal{A}_2 \sim z$. The goal of this section is to discuss how mirror symmetry is geometrically realised in the formulation presented in Section \ref{section2xx}.

As in the rest of this work, we restrict to {\it  balanced} quivers. Indeed, we discuss only the situation in which both the 
electric and the dual magnetic quivers are balanced. As we saw, this is possible only if there is only one flavour group, equivalently, only one stack of D5 branes, that 
have all the same linking numbers.
These conditions imply that the rank function of the electric and mirror magnetic quivers are `triangular'.

Consider then, the (electric) quiver field theory described by a generic rank function studied in Section \ref{generictriangular}.  
We summarise it in eq.(\ref{figureXX2}) to ease the reading.

\begin{equation}
    \begin{array}{c}
         \begin{scriptsize}
             \begin{tikzpicture}
                      \draw[->](0,0)--(4.5,0) node[right]{$\eta$};
                      \draw[->](0,0)--(0,3) node[above]{$\mathcal{R}_e(\eta)$};
                      \draw[scale=0.5, domain=0:4, smooth, variable=\x, blue] plot ({\x}, {\x});
                      \draw[scale=0.5, domain=4:8, smooth, variable=\x, blue] plot ({\x}, {-\x+8});
                      \draw[dashed, domain=0:2, smooth, variable=\y, gray] plot ({\y}, {2});
                      \draw[dashed,gray] (2,0) -- (2,2);
                      \node[label=left:{$SN$}]at(0,2){};
                      \node[label=below:{$S$}]at(2,0){};
                      \node[label=below:{$P$}]at(4,0){};
             \end{tikzpicture}
         \end{scriptsize}
    \end{array}\;;\quad \mathcal{R}_e(\eta)=\left\{\begin{array}{cc}
         N\eta & \eta\in[0,S]  \\
        \frac{SN}{P-S}(P-\eta)& \eta\in[S,P]
    \end{array}\right.\label{figureXX2}
\end{equation}
Let us summarise some numbers (number of branes, vectors, hypers and dimension of the Higgs branch) characterising the electric description of this quiver (see also figure \ref{quiver1+HW1}).
\begin{eqnarray}
& & N_{NS5}^{(e)}= P,\;\; N_{D5}^{(e)}={\cal R}_e'(0) -{\cal R}'(P)= \frac{P N}{P-S},\;\;\; N_{D3}^{(e)}=\int_0^P {\cal R}_e d\eta= \frac{N P S}{2}.\label{numberselectric}\\
& & n_{v}^{(e)} =\sum_{k=1}^S (N k)^2 +\sum_{k=1}^{P-S-1} \left(\frac{S N}{P-S} (P-S-k)\right)^2=\frac{N^2 P S}{6P-6S}(1+2 S P -2 S^2),\nonumber\\
& & n_{h}^{(e)} =\sum_{k=1}^{S-1} N^2 k (k+1) +\sum_{k=0}^{P-S-1} \left(\frac{S N}{P-S} \right)^2 (P-S-k)(P-S-k-1) + \frac{S P N^2}{(P-S)}=\nonumber\\
& & \frac{N^2 P S}{3P-3S}(2+ S P -2 S^2),\nonumber\\
& & \text{dim}~ {\cal M}_{H}^{(e)}=n_h^{(e)} - n_v^{(e)}= \frac{N^2 P S}{2P-2S}.\nonumber
\end{eqnarray}
Following the usual rules to construct the mirror dual, we exchange the linking numbers calculated  in eq.(\ref{linkingsex1}). The mirror system has,
\begin{eqnarray}
& & \hat{L}_{NS5_1}=\hat{L}_{NS5_2}=....= \hat{L}_{NS5_{\frac{PN}{P-S}  }}= P-S,\;\;\; ~~\hat{\rho}= [(P-S)^{ \frac{PN}{P-S}    }].\nonumber\\
& & L_{D5_1}= L_{D5_2}=....=L_{D5_P}= N,\;\;\;~~~~~~ \rho=([N]^P).\nonumber
\end{eqnarray}
The mirror system is encoded in the rank function and quiver in eq.(\ref{figureXX3}). 
\begin{equation}
\begin{array}{c}
\begin{scriptsize}
\begin{tikzpicture}
\node[label=above:{$P-S$}][u](1){};
\node[label=below:{$2(P-S)$}][u](2)[right of=1]{};
\node(dots)[right of=2]{$\cdots$};
\node[label=above:{$S(N+1)-P$}][u](3)[right of=dots]{};
\node[label=below:{$SN$}][u](mid)[right of=3]{};
\node[label=above:{$S(N-1)$}][u](3')[right of=mid]{};
\node(dots')[right of=3']{$\cdots$};
\node[label=below:{$S$}][u](1')[right of=dots']{};
\node[label=above:{$P$}][uf](f)[above of=mid]{};
\draw(1)--(2);
\draw(2)--(dots);
\draw(dots)--(3);
\draw(3)--(mid);
\draw(mid)--(f);
\draw(mid)--(3');
\draw(3')--(dots');
\draw(dots')--(1');
\end{tikzpicture}
\end{scriptsize}
\end{array}\;;\quad \mathcal{R}_m(\eta)=\left\{\begin{array}{cc}
(P-S)\eta & \eta\in[0,\frac{SN}{P-S}]  \\
S\left(\frac{NP}{P-S}-\eta\right)& \eta\in[\frac{SN}{P-S},\frac{NP}{P-S}]
\end{array}\right.\label{figureXX3}
\end{equation}
These electric and magnetic rank functions are generic under the restriction that
both quivers are balanced.
Let us now calculate the same numbers, using the magnetic quiver in eq.(\ref{figureXX3})
\begin{eqnarray}
& & N_{NS5}^{(m)}= \frac{P N}{P-S},\;\; N_{D5}^{(m)}= P,\;\;\; N_{D3}^{(m)}=\int_0^{ \frac{P N}{P-S}   } R_m(\eta) d\eta =\frac{N^2 P S}{2(P-S)}.\label{numbersmagnetic}\\
& & n_{v}^{(m)} =\sum_{k=1}^{\frac{S N}{P-S}} (P-S)^2 k^2 +\sum_{k=1}^{N-1} \left( S (N-k)\right)^2=\frac{N P S}{6P-6S}(P-S +2 S N^2),\nonumber\\
& & n_{h}^{(m)} =\sum_{k=1}^{ \frac{S N}{P-S} -1}  (P-S)^2 k (k+1) +\sum_{k=0}^{N-2} S^2 (N-k) (N-k-1) + N S P=\nonumber\\
& & \frac{N P S}{3P-3S}(2P -2 S + 3 N^2),\nonumber\\
& & \text{dim}~ {\cal M}_{H}^{(m)}=n_h^{(m)} - n_v^{(m)}= \frac{N P S}{2}.\nonumber
\end{eqnarray}
As expected, the number of D5 and NS-five branes is exchanged. Interestingly, the dimension of the Higgs branch of the  
electric theory is calculated by the number of D3 branes in the magnetic theory, and viceversa.

We calculate the Fourier coefficient of the magnetic rank functions and compare this with the same quantity for the electric rank function in eq.(\ref{akex1}). 
We find,
\begin{equation}
a_k^{(m)}= \frac{ (P-S) }{ NP}\int_0^{\frac{NP}{P-S} } R_{m}(\eta) \sin \left( \frac{k\pi(P-S) \eta}{NP} \right) d\eta= \frac{NP^2 }{ (P-S)\pi^2 k^2 }\sin\left( \frac{k\pi S}{P}\right).\label{akm}
\end{equation}
In other words,
\begin{equation}
a_k^{(e)}=a_k^{(m)}.\nonumber
\end{equation}
This should not surprise, as the holographic central charge (or the Free Energy) should coincide in both descriptions, namely
\begin{equation}
c_{hol}^{(e)}=\frac{\pi}{8}\sum_{k=1}^\infty k \left( a_k^{(e)}\right)^2=\frac{\pi}{8}\sum_{k=1}^\infty k \left( a_k^{(m)}\right)^2=c_{hol}^{(m)}.\nonumber
\end{equation}
Notice that also implies the equality of $\hat{W}_e (\sigma,\eta)$ and of $\hat{V}_e(\sigma,\eta)$ with their magnetic counterparts. We  summarise these findings in eq.(\ref{figXX4}).
\begin{equation}
\begin{array}{ccc}
     N_{NS5} & \longleftrightarrow & N_{D5}^m \\
     N_{D5}^e &\longleftrightarrow & N_{NS5}\\
     N_{D3}^e &\longleftrightarrow & \text{dim} \, \mathcal{M}^m_{H}\\
     \text{dim} \, \mathcal{M}^e_H &\longleftrightarrow & N_{D3}^m\\
     a_k^e\;;c_{hol}^e & \longleftrightarrow & a_k^m\;;c_{hol}^m \\
\end{array}\label{figXX4}
\end{equation}

\subsection{A purely geometric formulation of mirror symmetry}
With the restriction of having balanced quivers, both in the electric and the magnetic descriptions, we can formulate mirror symmetry purely in geometrical terms
by observing an interesting scaling on the electric rank function. In fact,  we scale the coordinate $\eta$ and the intervals $[a,b]$ according to,
\begin{equation}
\eta\leftrightarrow \frac{N_{NS5}}{N_{D5}} \hat{\eta},\;\;\; [a,b]\leftrightarrow \Big[\frac{N_{D5}}{N_{NS5}} a ,\frac{N_{D5}}{N_{NS5}} b \Big].\label{scaling}
\end{equation} 
Analysing this scaling for the electric rank function written in eq.(\ref{figureXX2}), we find the magnetic rank function in eq.(\ref{figureXX3}). Similarly, we can check that
$a_k^{(e)}\leftrightarrow a_{k}^{(m)}$.

In other words, we could `ignore' the existence of mirror symmetry, consider the electric rank function and perform the scaling in eq.(\ref{scaling}). We recover the rank function and quiver for the second field theory.
The D5 and NS5 branes get exchanged and the dimension of the Higgs branch is calculated by the number of D3 branes of the transformed theory. As a bonus, it is easy to see that both quiver field theories have the same $V(\sigma,\eta)$ and holographic central charge. The scaling in eq.(\ref{scaling}) is simple and could be applied to other systems with similar description.

\subsection{The scaling for generic rank functions}
Let us study an interesting by-product of our picture of mirror symmetry.

Consider a generic rank function and apply the scaling in eq.(\ref{scaling}). This will generically {\it not} produce the mirror dual. In fact, generically the mirror of a quiver is an unbalanced quiver, which is not described with the formalism we developed in this work. In other words, for generic balanced quivers with `polygonal' (rather than triangular) rank function, the scaling in eq.(\ref{scaling}) generates another balanced quiver.
This corresponds to a new CFT in which the role of NS5 and D5 branes is exchanged. The dimension of the Higgs branch of one theory is {\it not} calculated by the number of D3 branes in the transformed theory. Interestingly, both CFTs will share the same holographic central charge. Of course, this result might be a peculiarity of the holographic description and fail when $1/N$ corrections are taken into account. Let us consider an example to illustrate this point.

Consider a particular case of the second example of quivers discussed in Section \ref{trapezoidalgeneric}.
 Choose $M=Q=1, S=P-2$. We have the quiver and rank function  in eq.(\ref{figura5XX}).

\begin{equation}
    \begin{array}{c}
         \begin{scriptsize}
             \begin{tikzpicture}
                      \node[label=below:{$N$}][u](1){};
                      \node[label=below:{$N$}][u](2)[right of=1]{};
                      \node(dots)[right of=2]{$\cdots$};
                      \node[label=below:{$N$}][u](3)[right of=dots]{};
                      \node[label=above:{$N$}][uf](F1)[above of=1]{};
                      \node[label=above:{$N$}][uf](F2)[above of=3]{};
                      \draw(1)--(F1);
                      \draw(1)--(2);
                      \draw(2)--(dots);
                      \draw(dots)--(3);
                      \draw(3)--(F2);
                      \draw [thick,decorate,decoration={brace,amplitude=6pt,mirror,raise=20 pt},xshift=0pt,yshift=10pt]
(1) -- (3)node [black,midway,xshift=0pt,yshift=-35pt] {
$P-1$ nodes};
             \end{tikzpicture}
         \end{scriptsize}
    \end{array}\;;\quad \mathcal{R}(\eta)=\left\{\begin{array}{cc}
    
         N\eta & \eta\in[0,1]  \\
         N & \eta\in[1,P-1]  \\
        N(P-\eta)& \eta\in[P-1,P]
    \end{array}\right.\label{figura5XX}
    \end{equation}

This implies the numbers,
\begin{eqnarray}
& & Q_{NS5}= P,\;\;\; Q_{D5}= 2N,\;\;\; Q_{D3}= N(P-1).\label{teoria1}\\
& & n_{v}= N^2(P-1),\;\;\; n_{h}= N^2 P,\;\;\; \text{dim}~{\cal M}_\text{H}= n_h-n_v= N^2.\nonumber\\
& & {a}_k=\frac{N P}{k^2\pi^2}\left[ \sin\left( \frac{k \pi }{P}\right) + \sin\left( \frac{k \pi (P-1)}{P}\right) \right]  , \nonumber\\
 & & {c}_{hol}= \frac{N^2P^2}{32\pi^2}\text{Re}~ \left(7\zeta(3)  - 4 \text{Li}_3(e^{\frac{2\pi i}{P}}) +4 \text{Li}_3(- e^{\frac{2\pi i}{P}}) \right).\nonumber
\end{eqnarray}
We perform the rescaling,
\begin{equation}
\eta\to \frac{P}{2N}\hat{\eta},\;\;\;\;\; [a,b]\to \Big[\frac{2N}{P}a, \frac{2N}{P} b \Big].\nonumber
\end{equation}
This generates a rank function and quiver depicted in eq.(\ref{figure5XX}).
    \begin{equation}\begin{gathered}
        \begin{array}{c}
             \begin{scriptsize}
                     \begin{tikzpicture}
                         \node[label=below:{$\frac{P}{2}$}][u](1){};
                         \node[label=below:{$P$}][u](2)[right of=1]{};
                         \node(dots)[right of=2]{$\cdots$};
                        \node[label=below:{$N-\frac{P}{2}$}][u](3)[right of=dots]{};
                        \node[label=below:{$N$}][u](N1)[right of=3]{};
                        \node[label=below:{$N$}][u](N2)[right of=N1]{};
                        \node(dots')[right of=N2]{$\cdots$};
                        \node[label=below:{$N$}][u](N3)[right of=dots']{};
                        \node[label=below:{$N-\frac{P}{2}$}][u](3')[right of=N3]{};
                        \node(dots'')[right of=3']{$\cdots$};
                        \node[label=below:{$\frac{P}{2}$}][u](1')[right of=dots'']{};
                        \node[label=above:{$\frac{P}{2}$}][uf](F1)[above of=N1]{};
                        \node[label=above:{$\frac{P}{2}$}][uf](F2)[above of=N3]{};
                        \draw(1)--(2);
                        \draw(2)--(dots);
                        \draw(dots)--(3);
                        \draw(3)--(N1);
                        \draw(N1)--(F1);
                        \draw(N1)--(N2);
                        \draw(N2)--(dots');
                        \draw(dots')--(N3);
                        \draw(N3)--(F2);
                        \draw(N3)--(3');
                        \draw(3')--(dots'');
                        \draw(dots'')--(1');
                        \draw [thick,decorate,decoration={brace,amplitude=6pt,mirror,raise=20 pt},xshift=0pt,yshift=10pt]
(N1) -- (N3)node [black,midway,xshift=0pt,yshift=-35pt] {
$P-1$ nodes};
                     \end{tikzpicture}
             \end{scriptsize}
        \end{array}\;;
        \\
        \hat{\mathcal{R}}\left(\hat{\eta}\right)=\left\{\begin{array}{cc}
            \frac{P}{2}\hat{\eta} & \hat{\eta}\in[0,\frac{2N}{P}] \\
            N & \hat{\eta}\in[\frac{2N}{P},\frac{2N}{P}(P-1)]\\
            \frac{P}{2}(2N-\hat{\eta}) & \hat{\eta}\in[\frac{2N}{P}(P-1),2N] \\
        \end{array}\right.
        \end{gathered}\label{figure5XX}
    \end{equation}

 In this theory we calculate,
\begin{eqnarray}
& & \hat{Q}_{NS5}= 2N,\;\;\;\; \hat{Q}_{D5}=P,\;\;\;\; \hat{N}_{D3}=\frac{2N^2}{P} (P-1) , \label{teoria2}\nonumber\\
 & & \hat{n}_v= 2 \sum_{k=1}^{2N/P -1} \left(N-\frac{kP}{2}\right)^2 + N^2\left(1 + 2 N\left(1-\frac{2}{P}\right)\right)= \frac{N}{6P}(P^2+4N^2(3 P-4)) ,    \nonumber\\
 & & \hat{n}_h= 2 \sum_{k=0}^{2N/P-2} \left(N-\frac{k P}{2}\right) \left(N-(k+1)\frac{P}{2}\right)  + NP=\frac{2N}{3P} (P^2+ 2 N^2),\nonumber\\
 & & \text{dim}~\hat{{\cal M}}_H=N^3\left(\frac{4}{P}-2 \right) +\frac{N P}{2}  ,\nonumber\\
 & & \hat{a}_k=\frac{N P}{k^2\pi^2}\left[ \sin\left( \frac{k \pi }{P}\right) + \sin\left( \frac{k \pi (P-1)}{P}\right) \right]  , \nonumber\\
 & &\hat{c}_{hol}= \frac{N^2P^2}{32\pi^2}\text{Re}~ \left(7\zeta(3)  - 4 \text{Li}_3(e^{\frac{2\pi i}{P}}) +4 \text{Li}_3(- e^{\frac{2\pi i}{P}}) \right).\nonumber
 \end{eqnarray}
 Hence, we have two different theories, with the same central charge. In the case $P=2$ both theories discussed above are mirror pair.
 
 We close this analysis here, below we present a summary and the conclusions of this paper.
 \section{Conclusions}\label{conclusions}
Let us start with a brief summary of the contents of this paper.

In  Section \ref{section2xx}, we present a holographic formulation of ${\cal N}=4$ $d=3$ SCFTs describing the IR fixed point of balanced linear quivers of gauge group 
$\Pi_{i=1}^{P-1} U(N_i)$ and flavour group $\Pi_{j=1}^{P-1} SU(N_{f,j})$. The type IIB configuration in eq.(\ref{background}) include the presence of NS, D3 and D5 branes. Importantly,
it is written in terms of a Potential function $V(\sigma,\eta) $ or equivalently $\hat{W}(\sigma,\eta)$. This function solves a Laplace partial differential equation.
This PDE should be supplemented with boundary and initial conditions, hence defining an electrostatic problem.

It is in these initial conditions that the `kinematical' data of the dual CFT is encoded.
In the case we focused here, the quivers are balanced and the initial condition can be easily given in terms of a `rank function' ${\cal R}(\eta)$. By quantising Page charges, we learn in Section \ref{Pagechargessection}
that the rank function must be a piecewise continuous and linear function. The values of ${\cal R}(\eta)$ at integer values of the coordinate must also be integer, as it is associated with the number of branes in the corresponding Hanany--Witten set-up.

Given a balanced linear quiver, we present a clear procedure to automatically write the dual Type IIB configuration.
In this way, this work moves forward the project of giving an {\it electrostatic} description of all half-BPS AdS$_D\times S^2$ spaces in dimensions $D=2,3,4,5,6,7$.
In some dimensions $D=4$ (the case of interest in this work) and in $D=6$, there is a pre-existent formulation in the bibliography, based on a coupled of holomorphic functions \cite{DHoker:2007hhe}, \cite{DHoker:2016ujz}. We have clarified the map between our formulation and  that of \cite{DHoker:2007hhe}.

Also in Section \ref{Pagechargessection}, we defined a quantity that counts the number of degrees of freedom of the QFT. This quantity is proportional to the Free Energy of the field theory on $S^3$. We refer to it as holographic central charge.
In Section \ref{examplessection}, we worked out a set of examples and analysed the behaviour of the holographic central charge of these examples and special limits thereof. We make clear the non-perturbative character of the result, typically involving Polylogarithmic functions of order three in the parameters of the field theory.

In Section \ref{sectionmirrorxx}, we pedagogically presented   various aspects of the QFTs with emphasis on Mirror symmetry. The way in which our holographic  backgrounds display Mirror symmetry is discussed. We presented the mirror mapping between two holographic field theories and display the exchange of NS5 and D5 branes, the exchange of the dimensions of the Higgs and Coulomb branches (represented by the number of D3 branes in the system), the equality of the central charge and background for both descriptions, etc. As a byproduct of this analysis we discuss an operation that given a balanced quiver produces a different balanced one with the same holographic central charge as the original one.

This work opens new and interesting avenues for research, here a non-exhaustive list:
\begin{itemize}
\item{It seems natural to attempt to understand how our formalism can be extended/adapted to {\it non-balanced} quivers.}
\item{Exploring other observables, for example Wilson loops in different representations and their behaviour under Mirror symmetry.}
\item{It would be interesting to present a holographic description of the QFT with special-unitary (rather than unitary) gauge groups. This may follow the ideas of  \cite{Collinucci:2020kdm}, probably using recent developments in higher form symmetries.}
\item{Developing an analog description for circular quivers, following the work of \cite{Assel:2012cj}.}
\item{This work furthers the project of finding electrostatic description for holographic duals to linear quivers that flow to SCFTs$_d$, with $d=1,2,3,4,5,6$. It seems natural to study the relations among moduli spaces of these different SCFTs$_d$ along the lines of \cite{Akhond:2020vhc, Akhond:2021knl, Akhond:2021ffo}}.
\item{The program of relating half-BPS AdS-solutions and SCFTs needs more work, particularly in the cases of AdS$_3$ and AdS$_2$. The methods developed here suggest new AdS$_2$ backgrounds that would be nice to study.  }
\end{itemize}
We hope to come back to these problems soon.
\section*{Acknowledgments:}
We wish to thank various colleagues for conversations and their opinions on these topics. Among these colleagues we mention: Federico Carta,  Lorenzo Coccia, S. Prem Kumar,  Yolanda Lozano, Leonardo Santilli, Lucas Schepers.
\\
We are supported by STFC grant ST/T000813/1.
 Carlos Ireneo Nunez would like to remember Diego A. Maradona.

\appendix
\section{The Type IIB equations of motion}\label{appendix1}
In this appendix we write explicitly the equations of motion that our backgrounds  in eq.(\ref{background}) satisfy.
The action is
\begin{eqnarray}
& & S_{IIB,st}= \int_{M_{10}} \sqrt{g} \left[ e^{-2\Phi} \left( R +4 (\partial \Phi)^2  -\frac{1}{12} H_3^2\right) -\frac{1}{2} \left( F_1^2 +\frac{1}{6}F_3^2 +\frac{1}{240} F_5^2\right)  \right] -\nonumber\\
& & \frac{1}{2} C_4\wedge H_3\wedge dC_2.\label{IIBaction}
\end{eqnarray}
where the field in the democratic formalism are in general defined as
\begin{eqnarray}
& & H_3= dB_2,\;\;\; F_1= dC_0,\;\;\; \tilde{F}_3= d C_2 - C_0 H_3,\;\;\; F_5= dC_4 - H_3\wedge C_2,\label{fluxes}\\
& & F_1=* F_9, \;\;\;\; F_7= -* F_3.\nonumber
\end{eqnarray}
In the specific case considered here we have $F_1 = 0$.
The equations of motion are,
\begin{eqnarray}
& & R +4 \nabla^2\Phi -4 (\partial \Phi)^2 -\frac{H_3^2}{12}=0,\label{eqsmotionIIB}\\
& & 4R_{\mu\nu}+ 8\partial_\mu \Phi\partial_\nu\Phi -H_{\mu \sigma \rho }H_\nu{}^{\sigma \rho}+2 F_\mu F_\nu +F_{\mu \sigma \rho}F_\nu{}^{\sigma \rho}+\frac{1}{24}F_{\mu \sigma_1 \dots \sigma_4}F_\nu{}^{\sigma_1 \dots \sigma_4}
-g_{\mu\nu}\Big(F_1^2 +\frac{F_3^2}{6} \Big)=0.\nonumber\\[2mm]
& & d\left( e^{-2\Phi} * H_3\right)=- F_5\wedge F_3 - F_1\wedge F_7,\nonumber\\[2mm]
& & dH_3=0,\;\;\; dF_1=0,\;\;\; d F_3 - H_3\wedge F_1=0,\;\;\; d F_5 -H_3\wedge F_3=0,\nonumber\\[2mm]
& &  d*F_1 + H_3 \wedge *F_3=0,\;\;\;\; d * F_3 + H_3\wedge F_5=0.\nonumber
\end{eqnarray}

\section{Map to DEGK}\label{mapDEGK}
The DEGK solutions \cite{DHoker:2007hhe}, are  defined in terms of complex functions respect to the variable $z$:
\begin{eqnarray}
& & ds_{10,st}^2= f_1(z, \bar{z})\Big[ds^2(\text{AdS}_4) + f_2(z,\bar{z}) d s^2 (S^2_1)+ f_3(z,\bar{z}) d s^2 (S^2_2)+ f_4(z,\bar{z})d z d \bar{z} \Big], \nonumber\\[2mm]
& &e^{-2\Phi}=f_5(z,\bar{z}), \;\;\;\; B_2=f_6(z,\bar{z}) \text{Vol}(S^2_1),\;\;\;\; C_2= f_7(z,\bar{z}) \text{Vol}(S^2_2),
\end{eqnarray}
where
\begin{eqnarray}
& & f_1 = 2 \sqrt{-\frac{N_2 }{W}}, \;\;\;\; f_2 = -\frac{h_1^2 W}{N_1}, \;\;\;\; f_3 = -\frac{h_2^2 W}{N_2},\;\;\;\; f_4 = -2\frac{W}{h_1 h_2} , \;\;\;\; f_5 = \frac{N_1}{N_2} \nonumber\\[2mm]
& & f_6 = 4 \frac{h_1^2 h_2 \text{Im}(\partial_z h_2 \partial_{\bar{z}} h_1)}{N_1}+2 h_2^D , \;\;\;\; f_7 = 4 \frac{h_1 h_2^2 \text{Im}(\partial_z h_2 \partial_{\bar{z}} h_1)}{N_2}-2 h_1^D . \label{eq:def-fi_holomorphic}
\end{eqnarray}
The five-form field is given by
\begin{equation}
F_5 = \text{Vol(AdS}_4) \wedge d f_8+ *( \text{Vol(AdS}_4) \wedge d f_8)
\end{equation}
where
\begin{equation}
f_8 = 4 \left( 6 \text{Re} (\mathcal{C})- 3 \mathcal{D} - 2\frac{h_1 h_2}{W} \text{Im}(\partial_z h_1 \partial_z h_2) \right) \, .
\end{equation}
All these functions are defined in terms of two holomorphic functions $\mathcal{A}_{1,2}(z)$, in particular $h_{1,2}$ and $h_{1,2}^D$ are the dual real harmonic functions
\begin{equation}
h_1 = 2 \text{Im} (\mathcal{A}_1) \, , \qquad h_1^D = 2 \text{Re} (\mathcal{A}_1) \, , \qquad h_2 = 2 \text{Re} (\mathcal{A}_2) \, , \qquad h_2^D = -2 \text{Im} (\mathcal{A}_2).
\end{equation}
Also, we have  the following definitions
\begin{equation}
W= \partial_z\partial_{\bar{z}} (h_1 h_2) \, , \quad N_i = 2 h_1 h_i |\partial_z h_i|^2-h_i^2 W \, , \quad \mathcal{D} = 2 \text{Re}(\mathcal{A}_1 \bar{\mathcal{A}}_2) \, , \quad \partial_z \mathcal{C} = \mathcal{A}_1 \partial_z \mathcal{A}_2 - \mathcal{A}_2 \partial_z \mathcal{A}_1 .
\end{equation}
One can check that the background is invariant under conformal transformations $z \to f(z)$, and in particular we are free to choose one of the holomorphic functions as a coordinate. The second holomorphic function can be defined in terms of an auxiliary harmonic function $\hat{V}(z,\bar{z})$ as follows
\begin{equation}
\mathcal{A}_1 = \pi \partial_z \hat{V} \, , \qquad \mathcal{A}_2= \frac{\pi}{8} z \, .\label{B7}
\end{equation}
In order to match these backgrounds with those in eq. \eqref{background} we set
\begin{equation}
z = \sigma - i \eta \, , \qquad \hat{V} = \sigma V \, .
\end{equation}
With these identifications we have
\begin{eqnarray}
& & h_1 = \pi \sigma \partial_\eta V \, , \qquad h_1^D = \pi \partial_\sigma (\sigma V) \, , \qquad h_2 = \frac{\pi}{4} \sigma \, , \qquad h_2^D = \frac{\pi}{4} \eta \, , \\[2mm]
& & W= \frac{\pi^2}{8} \partial_{\sigma}(\sigma \partial_\eta V) \, , \qquad N_1 = \frac{\pi^4}{8}\sigma^3 \partial_\eta V \Lambda \, , \qquad N_2 = - \frac{\pi^4}{128} \sigma^3 \partial^2_{\eta\sigma} V \, . 
\end{eqnarray}
These expressions are enough to match \eqref{eq:def-fi_holomorphic} with eq.\eqref{background}.

\section{Useful expressions}\label{useful}
Let us write here some  expressions useful for the calculations in the main body of this paper.
We start with the translation between expressions containing $V(\sigma,\eta)$ in terms of $\hat{V}(\sigma,\eta)$.
\begin{eqnarray}
& & V(\sigma,\eta)=\frac{\hat{V}(\sigma,\eta) }{\sigma},\;\;\;\; \partial_\sigma V
=\frac{\partial_\sigma \hat{V} }{\sigma}-\frac{\hat{V} }{\sigma^2},\label{translation}\\
& & \partial^2_\sigma V= \frac{2 \hat{V} -2 \sigma \partial_\sigma \hat{V } +\sigma^2 \partial^2_\sigma \hat{V} }{\sigma^3},\;\;\;\; \partial_\eta V=\frac{\partial_\eta \hat{V} }{\sigma},\nonumber\\
& & \partial^2_\eta V=\frac{\partial^2_\eta \hat{V} }{\sigma},\;\;\;\; \partial^2_{\sigma\eta} V= \frac{\sigma \partial^2_{\sigma \eta}\hat{V} -\partial_\eta \hat{V} }{\sigma^2},\;\;\;\partial_\sigma\left(\sigma\partial_\eta V \right)=\partial^2_{\sigma \eta}\hat{V},\nonumber\\
& & \Lambda= \frac{\sigma (\partial^2_\eta \hat{V})^2 -\partial_\eta \hat{V} \partial^2_{\sigma \eta} \hat{V} +\sigma ( \partial^2_{\sigma \eta} \hat{V})^2 }{\sigma^2}.\nonumber
\end{eqnarray}
The various derivatives of $\hat{V}$ read,
\begin{eqnarray}
& & \hat{V}(\sigma,\eta)=
\sum_{k=1}^\infty a_k \cos\left( \frac{k\pi \eta}{P}\right) e^{-\frac{k\pi|\sigma |}{P}},\label{Vhatderivatives}\\
& & \partial_\eta \hat{V}=-
\sum_{k=1}^\infty a_k \left( \frac{k\pi}{P}\right)\sin\left( \frac{k\pi \eta}{P}\right) e^{-\frac{k\pi|\sigma |}{P}},\nonumber\\
& & 
\partial^2_\eta \hat{V}=-
\sum_{k=1}^\infty a_k \left( \frac{k^2\pi^2}{P^2}\right)\cos\left( \frac{k\pi \eta}{P}\right) e^{-\frac{k\pi|\sigma |}{P}},\nonumber\\
& & \partial_\sigma \hat{V}=-
\sum_{k=1}^\infty a_k \left( \frac{k\pi}{P}\right)\cos \left( \frac{k\pi \eta}{P}\right) e^{-\frac{k\pi|\sigma |}{P}} \text{sign}(\sigma),\nonumber\\
& & \partial^2_\sigma \hat{V}=
\sum_{k=1}^\infty a_k \cos \left( \frac{k\pi \eta}{P}\right) e^{-\frac{k\pi |\sigma |}{P}}\left( \frac{k^2\pi^2}{P^2} -\frac{k\pi}{P} \delta(\sigma)\right) ,\nonumber\\
& & \partial^2_{\sigma \eta} \hat{V}=
\sum_{k=1}^\infty a_k \left( \frac{k^2\pi^2}{P^2}\right)\sin \left( \frac{k\pi \eta}{P}\right) e^{-\frac{k\pi|\sigma |}{P}} \text{sign}(\sigma) \, .\nonumber
\end{eqnarray}
Using eqs.(\ref{translation})-(\ref{Vhatderivatives}), we find the following expressions,
\begin{eqnarray}
& & \frac{4}{\pi^2}f_1^2= \frac{\sigma}{\partial^2_{\sigma \eta} \hat{V}}\left(\sigma  \partial^2_{\sigma \eta} \hat{V}-\partial_\eta \hat{V} \right)=|\sigma|\frac{\sum_{k=1}^\infty a_k (\frac{k \pi}{P}) (1+\frac{k \pi |\sigma|}{P})     \sin\left( \frac{k \pi\eta}{P}\right) e^{-\frac{k \pi |\sigma|}{P} }     }{ \sum_{l=1}^\infty a_l \left(\frac{l^2\pi^2 }{P^2} \right) \sin\left( \frac{l \pi\eta}{P}\right) e^{-\frac{l \pi |\sigma|}{P}}    }.\label{f1}
\end{eqnarray}
For the combination $\sigma^2\Lambda$ we find,
\begin{eqnarray}
& & \sigma^2\Lambda= \sigma\left( \sum_{k=1}^\infty a_k \frac{k^2\pi^2}{P^2}     \cos\left( \frac{k \pi\eta}{P}\right) e^{-\frac{k \pi |\sigma|}{P} }      \right)^2 
+\sigma \left(   \sum_{k=1}^\infty a_k \frac{k^2\pi^2}{P^2}     \sin\left( \frac{k \pi\eta}{P}\right) e^{-\frac{k \pi |\sigma|}{P}  }       \right)^2 +\nonumber\\
& & \sum_{k,l=1}^\infty a_k a_l \frac{k^2\pi^2}{P^2} \frac{l \pi }{P}    \sin \left( \frac{k \pi\eta}{P}\right)  \sin \left( \frac{l \pi\eta}{P}\right)  e^{-\frac{(k+l) \pi |\sigma|}{P}} \text{sign}(\sigma).\label{Lambda}
\end{eqnarray}
Using this, we find for $f_2(\sigma,\eta)$
\begin{eqnarray}
& & f_2= -\frac{(\partial_\eta \hat{V} ) (\partial^2_{\sigma \eta} \hat{V})}{\sigma^2\Lambda}= \frac{1}{\sigma^2\Lambda}\left(   \sum_{k,l=1}^\infty a_k a_l \frac{l^2 k \pi^3 }{P^3}    \sin \left( \frac{k \pi\eta}{P}\right)  \sin \left( \frac{l \pi\eta}{P}\right)  e^{-\frac{(k+l) \pi |\sigma|}{P}} \text{sign}(\sigma)  \right).\nonumber
\end{eqnarray}
For the combinations $f_3(\sigma,\eta)$ and $f_4(\sigma,\eta)$ we find,
\begin{eqnarray}
& & f_3= 1+\frac{\partial_\eta \hat{V}}{\sigma \partial^2_{\sigma \eta} \hat{V} -\partial_\eta \hat{V}}=\frac{|\sigma| \sum_{n=1}^\infty a_n (\frac{n^2 \pi^2}{P^2})     \sin\left( \frac{n \pi\eta}{P}\right) e^{-\frac{n \pi |\sigma|}{P} }       }{ \sum_{k=1}^\infty a_k (\frac{k \pi}{P}) (1+\frac{k \pi |\sigma |}{P})     \sin\left( \frac{k \pi\eta}{P}\right) e^{-\frac{k \pi |\sigma|}{P} }       }.\nonumber\\
& & f_4= -\frac{\partial^2_{\sigma \eta} \hat{V}}{\sigma \partial_\eta \hat{V}}=\frac{\sum_{n=1}^\infty a_n (\frac{n^2 \pi^2}{P^2})     \sin\left( \frac{n \pi\eta}{P}\right) e^{-\frac{n \pi |\sigma|}{P} }     }{   \sum_{k=1}^\infty a_k (\frac{k \pi}{P})     \sin\left( \frac{k \pi\eta}{P}\right) e^{-\frac{k \pi |\sigma|}{P} } |\sigma|       }.\label{f3f4}
\end{eqnarray}
Finally, for the dilaton $e^{-2\Phi}= f_5(\sigma,\eta)$ we find,
\begin{eqnarray}
& &f_5= - \frac{16\sigma \Lambda \partial_\eta\hat{V}}{ \sigma^2\partial^2_{\sigma \eta} \hat{V} -\sigma \partial_\eta \hat{V}}=16\sigma\Lambda\left( \frac{     \sum_{n=1}^\infty a_n (\frac{n \pi}{P})     \sin\left( \frac{n \pi\eta}{P}\right) e^{-\frac{n \pi |\sigma|}{P} }          }{ \sum_{k=1}^\infty a_k (\frac{k \pi}{P}) (1+\frac{k \pi |\sigma|}{P})     \sin\left( \frac{k \pi\eta}{P}\right) e^{-\frac{k \pi |\sigma|}{P} }             } \right).\label{f5}
\end{eqnarray}
The asymptotic analysis of the backgrounds requieres the expressions
\begin{eqnarray}
& & f_1(\sigma\to\pm\infty,\eta)\sim \frac{\pi |\sigma|}{2},\;\;\; \sigma^2\Lambda\sim |\sigma| \frac{a_1^2 \pi^4}{P^4} e^{-\frac{2\pi |\sigma|}{P}},\label{asymptsigma}\\
& & f_2(\sigma\to\pm\infty,\eta)\sim \frac{P}{\pi|\sigma|}\sin^2\left(\frac{\pi \eta}{P}\right),\;\;\;\; f_3(\sigma\to\pm\infty,\eta)\sim 1,\nonumber\\
& & f_4(\sigma\to\pm\infty,\eta)\sim \frac{\pi}{P |\sigma|},\;\;\;\; f_5(\sigma\to\pm\infty,\eta)\sim \frac{e^{-\frac{2\pi |\sigma|}{P}}}{|\sigma|}.\nonumber
\end{eqnarray}
For the analysis close to $\eta\sim 0$, we use the expressions
\begin{eqnarray}
& &\frac{4}{\pi^2} f_1^2(\sigma,0)\sim |\sigma|
\frac{  \sum_{k=1}^\infty a_k (\frac{k\pi}{P})^2 (1+\frac{k\pi |\sigma | }{P} )  e^{-\frac{k\pi|\sigma |}{P}}    }{  \sum_{n=1}^\infty a_n (\frac{n\pi}{P})^3  e^{-\frac{n \pi |\sigma |}{P}}      }\label{asympeta}\\
& & \sigma^2\Lambda\sim \sigma  \left(  \sum_{k=1}^\infty a_k \left(\frac{k\pi}{P}\right)^2  e^{-\frac{k\pi|\sigma |}{P}} \right)^2,\;\;\; f_2(\sigma,0)\sim \frac{\eta^2}{|\sigma |} \left(  \frac{      \sum_{n,l=1}^\infty a_n a_l (\frac{n^2 l^3 \pi^5}{P^5})  e^{-\frac{(n+l)\pi|\sigma |}{P}}      }{      \left(  \sum_{k=1}^\infty a_k (\frac{k\pi}{P})^2  e^{-\frac{k\pi|\sigma |}{P}} \right)^2      }\right),\nonumber\\
& & f_3(\sigma,0)\sim \frac{ |\sigma|  \sum_{k=1}^\infty a_k (\frac{k\pi}{P})^3  e^{-\frac{k\pi|\sigma |}{P}}  }{        \sum_{n=1}^\infty a_n (\frac{n\pi}{P})^2 (1+\frac{n\pi \text{sign}(\sigma) }{P} )  e^{-\frac{n \pi|\sigma |}{P}}          }\nonumber\\
& & f_4(\sigma,0)\sim \frac{         \sum_{k=1}^\infty a_k (\frac{k\pi}{P})^3  e^{-\frac{k\pi|\sigma |}{P}}            }{  |\sigma |        \sum_{n=1}^\infty a_n (\frac{n\pi}{P})^2  e^{-\frac{n\pi|\sigma |}{P}}              },\;\;\;f_5(\sigma,0)\sim
16\frac{   \left(    \sum_{k=1}^\infty a_k (\frac{k\pi}{P})^2  e^{-\frac{k\pi|\sigma |}{P}}    \right)^2           }{        \sum_{n=1}^\infty a_n (\frac{n\pi}{P})^2 (1+\frac{n\pi |\sigma |}{P} )  e^{-\frac{n \pi|\sigma |}{P}}        }.\nonumber
\end{eqnarray}
Notice that, if we impose \eqref{eq:Ransatz}, we have the following expressions in the $\sigma \to 0$ limit:
\begin{equation}
f_1^2 = \frac{P^2}{4} \frac{\sum_k k ^2 a_k}{\sum_k k^4 a_k} \, , \qquad f_2 = \left(\frac{\pi \eta}{2 f_1}\right)^2  \, , \qquad f_3 = \left(\frac{\pi \sigma}{2 f_1}\right)^2 \, , \qquad f_4 = \left(\frac{\pi}{2 f_1}\right)^2 \, . \label{eq:asympetasigma}
\end{equation}

For the computation of Page charges, it is useful to calculate,
\begin{eqnarray}
& & \frac{\sigma \partial_\eta V \partial^2_\eta V}{\Lambda}
= \frac{ \sigma \partial_\eta \hat{V} \partial^2_\eta \hat{V}}{\sigma^2\Lambda}\sim_{|\sigma|\to\infty} \frac{P}{\pi} \sin\left(  \frac{\pi \eta}{P}\right) \cos\left( \frac{\pi \eta}{P}\right).\label{forB2}
\end{eqnarray}
The quantity ${\cal M}_1$ is,
\begin{eqnarray}
& & {\cal M}_1= \frac{\eta\sigma (\partial^2_\eta V)(\partial_\eta V) - \sigma(\partial_\eta V)^2}{\partial^2_{\sigma \eta}V}=\frac{\sigma       \left(    \eta  (\partial^2_\eta \hat{V}) (\partial_\eta \hat{V}) -(\partial_\eta \hat{V})^2      \right)
               }{\sigma \partial^2_{\sigma \eta}\hat{V}    -\partial_\eta \hat{V}       },\label{M1}\\
               & & {\cal M}_1=\frac{\sigma}{   \sum_{k=1}^\infty a_k \sin \left( \frac{k\pi \eta}{P}\right) e^{-\frac{k\pi |\sigma |}{P}}\left( \frac{k^2\pi^2 |\sigma|}{P^2} +\frac{k\pi}{P} \right)      }\times\nonumber\\
               & & \left[   \sum_{n,l=1}^\infty a_n a_l \eta  (\frac{l n^2 \pi^3 }{P^3})    \sin \left( \frac{l \pi\eta}{P}\right)  \cos\left( \frac{n \pi\eta}{P}\right)  e^{-\frac{(n+l) \pi |\sigma|}{P}}  - \left(         \sum_{r=1}^\infty a_r (\frac{r^2\pi^2}{P^2})     \cos\left( \frac{r \pi\eta}{P}\right) e^{-\frac{r \pi |\sigma|}{P} }          \right)^2    \right].\nonumber
 \end{eqnarray}
For the calculation of D3 Page charge we use the expression above to find
\begin{equation}
{\cal M}_1(\sigma\to+\infty,\eta)={\cal M}_1(\sigma=0^+,\eta)=0.\label{M1asymp}
\end{equation}
For  the quantity ${\cal M}_2$ we have
\begin{eqnarray}
& & {\cal M}_2= \partial_\sigma \left( \hat{W} -(\eta-\Delta)\partial_\eta \hat{W} \right)=\label{M2}\\
& &-  \sum_{k=1}^\infty a_k e^{-\frac{k\pi |\sigma |}{P}} \text{sign}(\sigma) \left[   \sin \left( \frac{k\pi \eta}{P}\right)  -(\eta-\Delta) (\frac{k\pi}{P})  \cos \left( \frac{k\pi \eta}{P}\right)  \right].\nonumber\\
& & {\cal M}_2 (\sigma\to +\infty,\eta)=0,\;\;\;\; 2{\cal M}_2(\sigma=0^+,\eta) =- R(\eta)+ (\eta-\Delta)R'(\eta).\nonumber
\end{eqnarray}
In the last step we have used eq.(\ref{rankfunction}).

For the calculation of the charge of D5 branes, we need the expression for
\begin{eqnarray}
& & \frac{f_7(\sigma,\eta)}{2\pi}=
-\partial_\sigma\hat{V} +\frac{    \sigma (\partial_\eta \hat{V} )(\partial^2_\eta \hat{V})     }{\sigma \partial^2_{\sigma \eta}\hat{V} -\partial_\eta\hat{V}     }=\label{f7}\\
& &  \sum_{k=1}^\infty a_k \cos \left( \frac{k\pi \eta}{P}\right) e^{-\frac{k\pi |\sigma |}{P}}\left(\frac{k\pi}{P} \right)  +
 \frac{  \sigma   \sum_{n,l=1}^\infty a_n a_l \eta  (\frac{l^2 n \pi^3 }{P^3})    \sin \left( \frac{n \pi\eta}{P}\right)  \cos\left( \frac{l \pi\eta}{P}\right)  e^{-\frac{(n+l) \pi |\sigma|}{P}}            }{       \sum_{k=1}^\infty a_k \sin \left( \frac{k\pi \eta}{P}\right) e^{-\frac{k\pi |\sigma |}{P}}\left( \frac{k^2\pi^2 |\sigma|}{P^2} +\frac{k\pi}{P} \right)               }.\nonumber\\
& & \frac{f_7(0^+,\eta)}{2\pi}=  \sum_{k=1}^\infty a_k (\frac{k\pi}{P})\cos \left( \frac{k\pi \eta}{P}\right)=\frac{1}{2}{\cal R}'(\eta).\nonumber
\end{eqnarray}
\section{The holographic central charge for a rank function with off-sets}\label{chol-offset}
We discuss an interesting generalisation to the result of Section \ref{centralgenericbalanced}. In particular, we discuss the situation for which $N_0$ and $N_P$ are nonzero. All the formal steps are  very similar, but in  this case, the Fourier coefficient in eq.(\ref{genericak}) reads,
\begin{eqnarray}
a_k= \frac{\left(  N_0 + (-1)^{k+1}  N_P \right)}{k \pi}\left[ 1-\frac{P}{k\pi}  \sin\left(\frac{k\pi}{P} \right)  \right] +\frac{P^2}{k^2 \pi^2} \sum_{j=0}^{P-1}F_j\sin\left(\frac{k\pi j}{P}\right).
\end{eqnarray}
Using now the expression for the holographic central charge (\ref{chol}) we find three terms,
\begin{eqnarray}
& & c_{hol}=\frac{\pi}{8}
\sum_{k=1}^\infty k a_k^2= T_1+T_2+T_3,\label{choloffset}\\
& &T_1=  \frac{P^2}{8 \pi^3}\sum_{k=1}^\infty\sum_{j,l=0}^{P-1} \frac{F_j F_l}{k^3}\sin\left(\frac{k\pi j}{P}\right)\sin\left(\frac{k\pi l}{P}\right) ,\nonumber\\
& & T_2= \frac{P}{4\pi^2} \sum_{j=1}^{P-1} F_j \sum_{k=1}^\infty \frac{    \left(  N_0 + (-1)^{k+1}  N_P \right)    \left( 1-\frac{P}{k\pi}  \sin\left(\frac{k\pi}{P} \right)  \right)  \sin\left(\frac{k \pi j}{P} \right)               }{k^2}    ,\nonumber\\
& & T_3= \frac{1}{8\pi}\sum_{k=1}^\infty\frac{     \left(  N_0 + (-1)^{k+1}  N_P \right) ^2   \left( 1-\frac{P}{k\pi}  \sin\left(\frac{k\pi}{P} \right)  \right)^2      }{ k           }  .\nonumber
\end{eqnarray}
The first term is the one we encountered in eq.(\ref{chol general balanced}). The term $T_2$ can be summed and we quote its final result
\begin{eqnarray}
& & T_2= \frac{P}{4\pi^2} \sum_{j=1}^{P-1} F_j \Big[ N_P \text{Im}\left( \text{Li}_2(- e^{-i \frac{\pi j}{P}}) \right) + N_0      \text{Im}\left( \text{Li}_2( e^{-i \frac{\pi j}{P}}) \right)  \nonumber\\
& & +\frac{N_P P}{2\pi} \text{Re} \left( \text{Li}_3 (- e^{i\frac{\pi (j-1)}{P}}) -    \text{Li}_3 (- e^{i\frac{\pi (j+1)}{P}})   \right) + \frac{N_0 P}{2\pi}   \text{Re} \left( \text{Li}_3 ( e^{i\frac{\pi (j+1)}{P}}) -    \text{Li}_3 ( e^{i\frac{\pi (j-1)}{P}})   \right)    
                      \Big] \, .\nonumber
\end{eqnarray}
 The term $T_3$ is curious, as it  contains the divergent harmonic sum ($\sum_{n=1}^\infty \frac{1}{n}$). Notice that this divergent term appears in previous gravity and CFT computations \cite{Assel:2011xz,Coccia:2020wtk,Coccia:2020cku}, so we need to regulate it. More explicitly, $T_3$ reads
 \begin{eqnarray}
 & & T_3=\frac{P^2}{32 \pi^3} (2 N_0^2+ 2 N_P^2 + 3 N_0 N_P)\zeta(3) +\frac{P N_0 N_P}{32\pi^3} \left[ 16\pi \text{Im}[\text{Li}_2(- e^{-i \frac{\pi}{P}}) ] - 2 P \text{\text{Re}} [\text{Li}_3(- e^{\frac{2\pi i}{P}})]  \right]\nonumber\\
 & &- \frac{P (N_0^2 +N_P^2)}{32\pi^3} \left[ 8\pi \text{\text{Im}[}\text{Li}_2(e^{i \frac{\pi}{P}}) ] +2 P \text{\text{Re}} [\text{Li}_3( e^{\frac{2\pi i}{P}})]  \right]\nonumber\\
 & & + \frac{(N_0^2+N_P^2)}{8\pi} \sum_{k=1}^\infty \frac{1}{k}  - \frac{N_0 N_P}{4\pi}\sum_{k=1}^\infty \frac{(-1)^k}{k}.\label{T3}
 \end{eqnarray}
The divergent sums in the last line can be regulated using the Ramanujan sum. This is a consistent fashion of extracting a finite part out of a divergent sequence.
Roughly the idea is to regulate
\begin{eqnarray}
& & \sum_{k=1}^\infty \frac{1}{k}~~\leftrightarrow ~~\lim_{N\to\infty} \left[ \sum_{k=1}^N \frac{1}{k} -\int_1^N \frac{dx}{x}  \right]= \gamma_E,\label{Ramanujan}\\
& &-\sum_{k=1}^\infty \frac{(-1)^k}{k}= -\sum_{n=1}^\infty \frac{1}{2 n} + \sum_{n=0}^\infty \frac{1}{2 n+1}~~\leftrightarrow~~ -\frac{\gamma_E}{2} + \lim_{N\to\infty} \left[ \sum_{n=0}^N \frac{1}{n} 
-\int_0^N \frac{dx}{2x +1}  \right]=\frac{\log 2}{2}.\nonumber
\end{eqnarray}
Where $\gamma_E$ is the Euler-Mascheroni constant. In this way we regulate $T_3$ and find a result for $c_{hol}$ in the case with offsets. Notice that in the holographic limit, when $P$ is very large
the regulated terms in $T_3$ are subleading.

\end{document}